\documentclass[ ]{copernicus2}

\usepackage{color}
\usepackage[T1]{fontenc}

\begin{document}

\title{Poynting's theorem in magnetic turbulence 
}

\author[1]{R. A. Treumann\thanks{Visiting the International Space Science Institute, Bern, Switzerland\\ \\ \emph{Correspondence to}: R. A.Treumann (art@geophysik.uni-muenchen.de)}}
\author[2]{W. Baumjohann}

\affil[1]{Department of Geophysics and Environmental Sciences, Munich University, Munich, Germany}
\affil[2]{Space Research Institute, Austrian Academy of Sciences, Graz, Austria}

\runningtitle{Turbulent Poynting flux}

\runningauthor{R. A. Treumann and W. Baumjohann}


\received{ }
\revised{ }
\accepted{ }
\published{ }


\firstpage{1}

\maketitle

{\bf Abstract. --} The Poynting theorem is used to obtain an expression for the turbulent power-spectral density as function of frequency and wavenumber in low-frequency magnetic turbulence. No reference is made to Elsasser variables as is usually done in magnetohydrodynamic turbulence mixing mechanical and electromagnetic turbulence. We rather stay with an implicit form of the mechanical part of turbulence as suggested by electromagnetic theory in arbitrary media. All of mechanics and flows is included into a turbulent response function which by appropriate observations can be determined from knowledge of the turbulent fluctuation spectra. This approach is not guided by the wish of developing a complete theory of turbulence. It aims on the identification of the response function from observations as input into a theory which afterwards attempts its interpretation. Combination of both the magnetic and electric power spectral densities leads to a representation of the turbulent response function, i.e. the turbulent conductivity spectrum $\sigma_{\omega \vec{k}}$ as function of frequency $\omega$ and wavenumber $\vec{k}$. {It is given as the ratio of magnetic to electric power spectral densities in frequency and wavenumber space. This knowledge allows for formally writing down a turbulent dispersion relation. Power law inertial range spectra result in a power law turbulent conductivity spectrum.  These can be compared with observations in the solar wind. {Approximations, warnings and caveats are found in the Appendix. } 

\section{Foreword}
To avoid any misconceptions, it seems worth mentioning that the aim of the present paper is not to contribute to progress in turbulence theory. It merely aims at the practical question of what can (possibly) be learned about turbulence respectively what can be extracted on the physics of turbulence just from the observation of the spectral densities of the electromagnetic field alone. 

This is a rather modest purpose which does by no means compete with any of the extraordinarily sophisticated, presumably complete theories of turbulence of which both Sir W. Lamb and W. Heisenberg intended to ask the Lord for lifting the `mystery' to them after -- hopefully -- entering the paradise. 

It will turn out that, indeed, some valuable physical information can, in principle, be extracted from the electromagnetic power spectral densities alone. However, the available measurements are manifestly incomplete and insufficient for this purpose. It requires the mutually independent measurement of the fully electromagnetic spectral densities with same resolution on spatial and temporal scales; moreover, spatial scales must be determined independently from temporal scales, and the covariance of electric spectral densities must be restored. The last point requires the measurement of the velocity fluctuations on the same scales, which sets high demands on measurements of the distribution function or otherwise implies techniques like antennas, double probes, injection of particle beams (for obvious reasons -- identification, mobility, stability with respect to high frequency kinetic fluctuations etc. -- preferentially ion beams in low frequency turbulence) which independently determine the turbulent electric fields.  

\section{Introduction}
During the last few decades magnetic  -- or magnetohydrodynamic -- turbulence has come into the focus of intense investigation both observationally \citep[for reviews cf., e.g.,][]{goldstein1995}, numerically \citep[cf., as for an example,][]{zhou2004,dmitruk2009} as also theoretically \citep[cf., e.g.,][]{scheko2009,boldyr2013,zank2012} or semi-theoretically \citep[e.g.,][]{sahraoui2012}. A not anymore recent, though still valuable systematic recollection of some of its basic aspects, is found in the monography of \citet{biskamp2003}. For an earlier, now a bit outdated account, though including the development of chaotic motions and intermittency, see \citet{frisch1995}. 

Magnetic turbulence should not be mixed up with genuine plasma turbulence which is the development of turbulence in the kinetic regime where a large number of kinetic processes participate in generating high-frequency electromagnetic waves in the interaction between unstable particle distributions \citep[cf., e.g.,][and references therein]{davidson1972,kadomtsev1965,tsytovich1970,sagdeev1969}. Magnetic turbulence we have in mind is restricted to such low frequencies that it cannot leave the medium but is completely confined to the volume under consideration. It is not radiative and not electrostatic but consists of turbulent flows which mix up both an internally pre-existing magnetic field and current distribution which is stirred up by the turbulent motions and forms current vortices. These are surrounded by their solenoidal magnetic fields. There are no free electric charges under such conditions and at those low frequencies as these have plenty of time to discharge. Moreover, the turbulence in such media is collisionless under the conditions which interest us, for instance in space and in most astrophysical situations as well. This differs strongly from the interior of planets where turbulence is made responsible for the generation of the magnetic field via dynamo effects. 

The problem of turbulence splits into two different parts: the turbulent mechanical motion and its consequence and counterpart,  turbulence in the electromagnetic field. The mechanical motion of the turbulent medium affects the electromagnetic field via the transport equations. The conventional approach \citep[for a most recent theoretical treatment including transport see][]{zank2012} is based on a mixture of both parts. {This is guided by low frequency Alfv\'en waves which couple the velocity $\delta\vec{v}$ and magnetic $\delta\vec{B}$ fluctuation fields, and is done by introducing mixed fields $\vec{z}^\pm=\delta\vec{v}\pm\delta\vec{B}/\sqrt{\mu_0\rho}$, with $\rho$ mass density, the so-called Elsasser variables \citep[defined two thirds of a century ago by W.][]{elsasser1950}. They have the advantage of distinguishing between directions of propagation and, if combined with magnetic helicity, also between polarizations of the electromagnetic field. The mixture of mechanical and electromagnetic fields leads to a complicated set of transport equations for the turbulent correlation functions. Their currently probably most complete analytical versions including macroscopic viscosities are found in \citet{zank2012}, where correlation times and lengths have been expressed in terms of various ratios, and spectral powers have been listed. Weak turbulence has also been based on collisionless fully kinetic theory \citep[cf.,][]{yoon2007a,yoon2007b}. } 

Below we follow a different route concentrating on the electromagnetic part while only implicitly acounting for the mechanical part.  Turbulent energy is provided almost exclusively by mechanical motions, which is particularly true when the magnetic fields are comparably weak with kinetic energy density exceeding the magnetic energy density. This is the usual case in the solar wind, for instance and anywhere else in space and most extended astrophysical systems where magnetic fields are comparably weak, the order of $\mu$G to mG. In magnetically active media mechanical turbulence is even more difficult to treat than in the absence of magnetic fields, except in the most simple cases. At temperatures above ionisation temperature neglect of currents and fields is hardly justified anyway. 

Turbulence, as is well known, evolves in different stages. Firstly, the injection of energy takes {usually} place at the largest {spatial scales,} typical for mechanical motions {at or above ordinary sound wavelengths. Corresponding fluctuation frequencies are around the frequency of sound, magnetically being in the range of magnetosonic and Alfv\'enic waves. In a very simplified picture,  the injected electromagnetic energy flows across the spectrum of increasing wavenumbers $\vec{k}$, cascading down in scales until reaching the so-called microscopic or ``molecular scale'' where dissipation takes over. This is, in principle, the picture\footnote{{This picture is  a shorthand of the various processes which take place in the evolution of turbulence. Firstly, the injected energy is mostly mechanical, causing velocity and density perturbations, fluid vortices and all kinds of large-scale turbulent motions. In a conducting fluid or gas these lead to large-scale current fluctuations with their proper magnetic field fluctuations. By some badly known processes the large-scale currents and current vortices decay to form shorter scale current vortices and their magnetic fields. At the large scales these are in the magnetohydrodynamic/magnetogasdynamic domain. Once the scales drop below typical ion scales (gyroradii, inertial lengths), ions decouple magnetically but still respond to induction electric fields. Currents are transported by electrons here.} {This is the regime of Hall currents, kinetic Alfv\'en waves, whistlers etc., waves and instabilites whose magnetic fields are dominated by electrons. Current scales in this regime range  from just below the ion-gyroradii down to the electron gyroradius. At this stage the magnetohydrodynamic description breaks down, is replaced first by two-fluid processes, and ultimately by kinetic processes. The injected energy has already crossed several regimes of scales in the} {cascade from large to small scales. Transformation of wave modes has occurred, waves have generated sidebands, have undergone parametric interaction, refraction from large scale vortices, particle scattering and other nonlinear processes typical for each separate regime of scales have occurred, including various forms of dissipation, heating and entropy production. Nevertheless in this cross-scale process a regime might exist, where dissipation is about negligible. In this scale range energy cascades down to smaller scales to form a power law spectrum. Once its end is reached, which in magnetic turbulence is at electron scales, where the electrons effectively decouple from the magnetic field, still orders of magnitude above any `molecular' scales and the corresponding molecular dissipation, magnetic fluctuations cease, currents and magnetic fields dissipate, and magnetic turbulence is replaced by electrostatic turbulence. This is the regime of collisionless turbulent dissipation. There has been dissipation already at intermediate scales, but the bulk of dissipation takes place here by not well defined processes which destroy current flow and magnetic fields. The most violent of such processes is spontaneous magnetic reconnection \citep[cf., e.g.,][]{treumann2015} which sets on when the width of the turbulent current filaments drops below the electron gyroradius. {Dissipation in a single current filament is of course small. However, current filaments in turbulence are abundant, and one expects  that small-scale reconnection, integrated over the distribution of all current filaments in the turbulent volume, adds up to become} the ultimate kinetic dissipation mechanism in magnetic turbulence in plasmas, a kinetic process that not anymore belongs to magnetic turbulence, is dissipative in nature, and in which electrostatic fields play not only an important but the dominant role.} }} that has become accepted since the pioneering work of \citet{kolmogorov1941}, \citet{batchelor1947}, \citet{heisenberg1948a,heisenberg1948b}, \citet{iroshnikov1963}, \citet{kraichnan1964} and others. Meanwhile a large number of sophisticated theories have been attempted on the spectral behavior, {even including renormalization group methods \citep[cf., e.g.,][]{verma2004}.} Observations, mainly in the solar wind, provided strong support for Kolmogorov's hypothesis of the existence of an inertial spectral {subrange} between the injection and dissipation ranges {\citep[cf., e.g.,][and others]{goldstein1995,zhou2004, bale2005, alexandrova2009,sahraoui2009,sahraoui2013,chen2011,wicks2012,horbury2012,brown2015} with a fairly robust spectral power law}. They have been supported by numerical simulations under various collisionless conditions \citep[cf., e.g.,][and the extended lists of references therein]{zhou2004,bereznyak2011,bereznyak2014}. Reviewing these developments would fill a monography. 

We here focus on the question what can be deduced from the knowledge of the power spectral density of (electro)-magnetic turbulence in view of the electromagnetic response to the turbulent mechanical motion. {This requires investigation of the electromagnetic energy transport. Since, as mentioned above, the turbulent energy is provided by mechanical motion of which, due to the conducting properties of the plasma, just a fraction transforms into currents with their electromagnetic fields, energy conservation concerns the total sum of mechanical and electromagnetic energy. It is somewhat surprising then that observed electromagnetic spectral power densities exhibit a moderately extended spectral range which shows all signs of a} {Kolmogorov inertial range, characterized by a constant about dissipationless electromagnetic energy flow across the spectrum from large to small scales. In traversing it the electromagnetic energy is, at least approximately, conserved independently from mechanical energy, posing the purely electromagnetic problem of how the electromagnetic energy evolves over the scales of the inertial range. While most of the mechanical energy should probably have been dissipated by different, more violent mechanical processes, dissipation of the decoupled and  transferred electromagnetic energy is believed to take place in the small-scale dissipation range. }

Since turbulence consists of two states, the mean-field state and turbulent fluctuations, interest becomes divided. Mean fields are the sum of possibly present external (magnetic) fields and the fluctuation-averaged large-scale (electro)magnetic and mechanical fields. Both naturally affect the transformation of turbulent energy, mechanical and electromagnetic. Here we are interested in the electromagnetic part and, particularly, in the electromagnetic fluctuations which are measured in well-developed and stationary turbulence. 

{Our intention is as follows: Turbulence concerns fluctuations, including their nonlinear evolution to large amplitudes, mode coupling and mixing of scales. Observations refer to fluctuations when discussing power spectral densities of (electro)-magnetic turbulence. In the following we deal with the spectral energy density of these fluctuations. As these are the easiest available, we ask what information can be deduced from the electromagnetic power spectral densities alone when information about the mechanical part of turbulence is not at hand. Clearly, since both are coupled, such an approach cannot be completely general but will be restricted to a few special cases only where the mechanical aspect of the turbulence is assumed to be implicit to the electromagnetic response of the turbulent plasma. Under some restrictions, as we demonstrate below, this becomes possible for low frequency turbulence in application to the inertial range of turbulence.}

{The main restriction is that we consider only the electromagnetic contributions to turbulence. This implies that electrostatic fields are excluded. Thus $\nabla\times\delta\vec{E}=0$ and $\delta\vec{E}\perp\delta\vec{B}$ hold for the electromagnetic fluctuation fields. It is in this sense that we speak of magnetic turbulence. Its electromagnetic signature is the presense of a spectrum of fluctuating magnetic fields $\delta\vec{B}$ and the related induction electric fields $\nabla\times\delta\vec{E}=-\partial_t\delta\vec{B}$.}

{It is sometimes claimed that this would not cover turbulence, for instance in magnetohydrodynamics where Alfv\'en, fast and slow modes are involved and all fluctuations propagate on a combination of these low frequency modes while at higher frequencies change to two-fluid modes or electron-mhd modes with electric potential fields becoming gradually involved due to the charge separation effects caused by the different scales of electrons and ions.\footnote{{The argument that low frequency waves like Alfv\'en, fast and slow waves are not `real' electromagnetic waves, because they don't leave the plasma, differ from radiation and apparently ``include potential effects'', is wrong.  Suppression of radiation of Alfv\'en waves is due to mass loading, not to non-electromagnetic properties. This is seen from their dispersion relation $\mathcal{N}^2\equiv k^2c^2/\omega^2=c^2/V_A^2$ where $V_A^2=B^2/\mu_0mN$ is the squared Alfv\'en speed. For electromagnetic radiation the right-hand side is unity. Writing the Alfv\'en dispersion relation $\mathcal{N}^2=m/M^A_{\mathit{eff}}$, it becomes obvious that the radiative property of the electromagnetic  wave is lost, because the low-frequency photons assume an `effective Alfv\'en-mode mass' $M^A_{\mathit{eff}}=B^2/\mu_0Nc^2=\epsilon_0B^2/N$, which retards them to Alfv\'en speed. Photons become suddenly heavy and slow due to a `classical Higgs effect' in the presence of the magnetic field $B$ and plasma density $N$, which confines them to the plasma, because \emph{heavy photons don't propagate in vacuum}. In case this mass is fine-tuned to $M_{\mathit{eff}}^A= m$, Alfv\'en waves turn into ordinary-mode radiation of different dispersion and wavenumber-dependent effective ordinary-mode mass $M^\mathit{om}_{\mathit{eff}}=(m+e^2N/\epsilon_0k^2c^2)$. Note that for $k\to0$  this mass $M^\mathit{om}_{\mathit{eff}}\to\infty$ becomes infinite, and the ordinary mode is cut off at large wavelengths. It is thus the `Higgs-effect of mass generation' which transforms electromagnetic radiation in a magnetically active plasma into plasma waves. {The \emph{irreducible} ordinary-mode photon mass follows for $k=0$ from $m_\mathit{om}=\hbar\omega_e/c^2\approx 10^{-10}\sqrt{N}$ eV/$c^2 \approx 10^{-15} m_e\sqrt{N}$, where the density is in cm$^{-3}$. No irreducible mass exists for Alfv\'en modes because for them $\omega\to0$ with $k\to0$. Their effective irreducible mass is determined by the size of the system, $k_\mathit{min}$, the smallest wavenumber.}}} Such a claim is not true, however. Any fluctuating magnetic fields $\delta\vec{B}$ and their related induction electric fields $\delta\vec{E}$ are solely and without any exception due to the presence of electric currents on various different spatial and temporal scales. These are generated by whatsoever, however mostly mechanical mechanisms in the turbulence. Examples are density gradients which produce diamagnetic drift currents that give rise to magnetic fields and whose density fluctuations can be described as drift modes. They propagate on the fast or slow branches or their modifications at shorter scales. Electrostatic potential drops generate beams and thus currents whose magnetic fields contribute to magnetic turbulence in one of the electromagnetic modes, not the potentials themselves. The presence of magnetic fluctuations in the turbulence in a plasma thus always implies the presence of time variations on vastly different scales in the current distribution  -- no need to mention that spin couplings and orbital magnetic moments are excluded. On the other hand, potential fluctuations not related to currents do not contribute to magnetic fluctuations other than electromagnetic radiation.} 

\section{Turbulent magnetic energy transport equation}

Consider an infinitely extended magnetically active and collisionless plasma. {It has been shown \citep{boozer1986} that in the absence of any molecular collisions and kinetic effects mimicking collisions the only dissipation of energy in this case is due to turbulent fluctuations. In the following we derive an approximate equation for the dissipated energy from Poynting's law under some simplifying assumptions. These include consideration of fluctuation scales shorter than any mean field variations but longer than kinetic scales such that the fields can be divided into mean and fluctuating components. Kinetic scales are excluded because they are responsible for dissipation. Mean fields are assumed either vanishing (electric) or stationary (magnetic) such that the equations describe solely fluctuations. Moreover, velocity fluctuations are only accounted for implicitly where they contribute to the effective turbulent conductivity.} 

The electromagnetic energy flux in a plasma which carries a current $\vec{J}$ is given by Poynting's theorem \citep{landau1989}
\begin{equation}\label{eq-poynting}
\frac{\partial}{\partial t}\left(\frac{B^2}{2\mu_0}+\frac{\epsilon_0}{2}\vec{E'\cdot\tilde{\vec{\in}}\cdot E'}\right) =-\vec{J\cdot E'}-\frac{1}{\mu_0}\vec{\nabla\cdot\big(E'\times B\big)} 
\end{equation}
where $\tilde{\vec{\in}}$ is the tensor of dielectric response, and $\vec{S=E'\times B}/\mu_0$ is the Poynting flux vector. {One should keep in mind that this equation just desribes the energy in the electromagnetic field. In turbulence this field couples to the mechanical equations of motion of the plasma, exchanges energy with the plasma and thus is not conserved. Conservation of energy applies only to the sum of mechanical and electromagnetic energies. In exactly this sense the above equation is not a conservation law but applies to the \emph{open} electromagnetic system of turbulence under consideration! Energy losses and gains are contained in the interaction term $\vec{J\cdot E'}$ via the current $\vec{J}$ as well as in the electric field $\vec{E'}$ which both are functionals of the velocity $\vec{v}$.} 

In view of application to spectral densities of the turbulent fields, we assume that all field quantities $\vec{F}=\bar{\vec{F}}+\delta\vec{F}$ are composed of sums of fluctuation fields $\delta\vec{F}$ and mean fields $\bar{\vec{F}}$. The latter result from averaging over the timescales and spatial scales of the fluctuation fields as indicated by either the bar on top or $\langle\dots\rangle$.\footnote{{cf. Appendix  for the averaging procedure.}}  They are taken constant on the fluctuation scales thus not being of primary interest here. In addition large-scale external magnetic fields $\vec{B}_0$ may be present. Transforming into the moving-mean frame $\bar{\vec{v}}=0$, the mean electric field $\bar{\vec{E}}=0$ vanishes\footnote{{The electric field $\vec{E}'=\vec{E}+\vec{v}\times\vec{B}$ depends on the velocity field $\vec{v}$, which in a mean field theory gives rise to a modified Ohm's law including mean-field $\alpha$ and $\beta$ effects, with $\vec{E}'$ the electric field in the stationary frame. The electric fluctuations consist of three terms $\delta\vec{E}'=\delta\vec{E}+\delta\vec{v}\times\bar{\vec{B}}+\delta\vec{v\times}\delta\vec{B}$. No assumption is made about the velocity except that $\bar{\vec{v}}=0$. In the following the third nonlinear fluctuation term is neglected because it leads to higher order correlations. In the same spirit the average term $\overline{\delta\vec{v\times}\delta\vec{B}}$, the term leading to the $\alpha$-effect, is neglected, which also means that $\bar{\vec{E}}=0$.}} while magnetic mean fields $\bar{\vec{B}}$ have to be retained because Lorentz invariance keeps them unaffected. However, all averages of products of fluctuations like $\langle\delta\vec{F}_i\delta\vec{F}_j\cdots\rangle$ vary on the macroscopic scale. They provide constant contributions to the equations for fluctuations. In a first treatment they will be neglected. {If retained, the solutions obtained from the fluctuation equations without these terms must be adjusted to them by standard methods of variation of constants or reference to boundary conditions.} 

{Poynting's theorem results fom the set of electrodynamic equations. These are linear in the fields. In order to proceed without invoking the full set of mechanical equations whether fluid or kinetic, some assumption is to be made on the relation between the current $\vec{J}$ and electric fields, which includes the dissipative response of the medium to the turbulent fluctuations. This is accounted for through introducing a general conductivity tensor $\tilde{\vec{\sigma}}$ and a generalized Ohm's law $\vec{J}=\tilde{\vec{\sigma}}\cdot\vec{E}'$ in which the conductivity becomes a very complicated not further specified expression, depending on which plasma model is used to account for the turbulent dynamics \citep[cf., e.g.,][]{krall1973}. Avoiding any specification somehow circumvents the enormous difficulties involved in complete turbulence theory. This can be achieved only on the expense of a number of severe simplifications and assumptions to which we restrict. The present approach must therefore be considered as approximative only while in a similar sense it parallels the common experimental approach when just power spectral densities of the magnetic field are measured in order to infer spectral slopes.} 

Averaging Eq. (\ref{eq-poynting}) and subtracting the mean-field equation from the original Poynting theorem one obtains the Poynting theorem of the fluctuation fields alone. Understanding all terms as turbulent fluctuations we suppress the prefix $\delta$, but keep the bar-indication of the average (plus external) field
\begin{eqnarray}\label{eq-fluct-one}
\frac{\partial}{\partial t}\Big(\frac{B^2}{2\mu_0}+\frac{\vec{B\cdot}\vec{\bar{B}}}{\mu_0}&+&\frac{\epsilon_0}{2}\vec{E'\cdot\tilde{\vec{\in}}\cdot E'}\Big)=-\vec{E'\cdot\tilde{\sigma}\cdot E'}\nonumber\\[-1.5ex]
&&\\[-1.5ex]
&-&\frac{1}{\mu_0}\vec{\nabla\cdot\Big[E'\times B+E'\times\bar{B}\Big]}\nonumber
\end{eqnarray}
where $\tilde{\vec{\in}}=\bar{\vec{\in}}+\delta{\vec{\in}},\ \tilde{\vec\sigma}=\bar{\vec\sigma}+\delta\vec{\sigma}$. Any mean-field conductivity \citep{boozer1986} is included in $\bar{\vec{\sigma}}$ which contains only mean-field quantities. {\citet{boozer1986} for instance writes the \emph{mean-field resistivity tensor} in the form
\begin{displaymath}
\bar{\vec{\sigma}}^{-1}_\mathit{mf}= \bar{\eta}\, \textsf{I}\!\!\!\!\!\textsf{I}-\frac{\bar{\eta}_\|}{\bar{B}^2}\bar{\vec{B}}\bar{\vec{B}},\quad \bar{\eta}_\|=\frac{1}{\bar{B}}\nabla\cdot\Big(\lambda\nabla\frac{\bar{J}_\|}{\bar{B}}\Big)\nonumber
\end{displaymath}
with $\lambda>0$ some parameter that is constant on the fluctuation scale but variable on the mean-field scale, $\bar{J}_\|$ the mean-field aligned current density, and $\bar{\eta}$ an isotropic anomalous resistance resulting from kinetic theory, presumably a constant or slowly varying mean-field-scale quantity as well. The mean-field turbulent effect is thus contained only in $\lambda$ and the parallel current density $J_\|$ yielding $\bar{\eta}_\|$, an expression similar to the ordinary $\alpha$-effect. Since we are not interested in mean-field dynamics which becomes important in dynamo theory, we do not go further into any detail. Our interest focusses on the turbulent spectra of the fields. We simply assume that $\bar{\vec{\sigma}}_\mathit{mf}$ is part of our general time-dependent conductivity $\tilde{\vec{\sigma}}\big(t,\vec{x}(t)\big)$ which contains the response to the turbulent fluctuations. The $\delta$-transport terms contain their fluctuating parts. Since no model assumption will be made on them in the following, we do not distinguish between the mean and fluctuating dissipative components of the conductivity but refer to $\tilde{\vec{\sigma}}$ in their fluctuation-scale dependent combination.}

Through Ohm's law, the conductivity $\tilde{\vec{\sigma}}$ is a function of the field fluctuations plus mean fields and is thus modified by the turbulence. In contrast to other approaches \citep[cf., e.g.,][and others]{zank2012}, we do not assume any explicit model for its functional dependence. $\tilde{\vec{\sigma}}$ is taken as unknown in the attempt to express it solely through the observed fields respectively their energy densities.
Poynting's theorem for the fluctuations is genuinely nonlinear and general as no approximation has been made yet except for the averaging procedure {which restricts its validity to scales in space and time short with respect to the variations of the mean fields}.  All mechanical contributions are contained in the expression for the current density $\vec{J}$, i.e. the conductivity and the dielectric response. {In not making any reference to a particular model of the dynamics of matter, the fluid or plasma, no further assumption is made about these quantities, i.e. no assumption is made about the kind of mechanical nonlinearities which are thus retained in full in leaving $\tilde{\vec\sigma}$ unspecified.} 

Equation (\ref{eq-fluct-one}) contains two terms linear in the fluctuation fields whose presence complicates the discussion. They could, in principle, be retained. However, for the sake of simplicity we chose to drop them by the following arguments.
There are two ways of getting rid of them. Consider propagation of all fluctuations parallel to the mean magnetic field $\bar{\vec{B}}$ and take into account that the fluctuations are all solenoidal{, i.e. they are purely electromagnetic}. Then $\vec{E', B}$ and $\bar{\vec{B}}$ are all orthogonal, and the dot-products of the magnetic fluctuations and mean fields vanish. {Only the transverse component $\sigma^T$ of the conductivity is involved in this case. The divergence of the linear term in the Poynting flux can be written as $\vec{\bar{B}\cdot(\nabla\times E')}$. The term in the brackets is the negative time derivative of the magnetic fluctuation field. Its contribution to the divergence of the Poynting flux vanishes for parallel propagation, i.e. for $\vec{k}\|\vec{\bar{B}}$.} 

{The other way is to assume perpendicular propagation $\vec{k}\perp\bar{\vec{B}}$ with the electric fluctuation field $\vec{E}'\|\bar{\vec{B}}$ parallel to the mean magnetic field, which also implies that $\delta\vec{v}\times\bar{\vec{B}}=0$ in this case does not contribute in Ohm's law. Only the parallel conductivity $\sigma^\|$ is involved here.}

{In these two cases of propagation both linear terms disappear for physical reasons. One should, however, note that this holds only because the restriction is to low-frequency electromagnetic turbulence. This involves the absence of electric charges, in other words it excludes all kinds of dielectric responses which refer to the generation of electrostatic fluctuations. We will return to this restriction in the discussion of the meaning of the turbulent dissipation range.} 

In both {these cases the Poynting theorem} (\ref{eq-fluct-one}) for the turbulent fluctuations reduces to its original simplifyied form (\ref{eq-poynting}) which just holds for the fluctuations in the two special cases under consideration with respect to the mean magnetic field direction $\bar{\vec{B}}$ and excluding any compressive magnetic fluctuations\footnote{Clearly, consideration of only these two cases of propagation does not describe the full electromagnetic spectrum of turbulence. It lacks the missing third magnetic component (see the explanation below).}
\begin{equation}\label{eq-fluct-two}
\frac{\partial}{\partial t}\Big(\frac{B^2}{2\mu_0}+\frac{\epsilon_0}{2}\vec{E\cdot\tilde{\vec{\in}}\cdot E}\Big)=-\vec{E\cdot\tilde{\sigma}\cdot E}-\frac{1}{\mu_0}\vec{\nabla\cdot\Big(E\times B\Big)}
\end{equation}
where now all quantities are understood as fluctuations and the prime has been dropped. The first term on the right has the same structure as the second on the left. When brought to the left it can be incorporated as
\begin{eqnarray}
\frac{\partial}{\partial t}\bigg[\,\frac{B^2}{2\mu_0}+\frac{\epsilon_0}{2}\vec{E\cdot\tilde{\vec{\in}}\cdot E} &+&\!\!\!\!\int_{-\infty}^t\!\!\!\!\!\!\!\!\mathrm{d}t'\ \vec{E}(t')\ \vec{\cdot\ \tilde{\sigma}}(t')\ \vec{\cdot\ E}(t')\, \bigg] =\nonumber \\[-1.5ex]
&&\\[-1.5ex]
&=&-\frac{1}{\mu_0}\vec{\nabla\cdot\Big(E\times B\Big)}\nonumber
\end{eqnarray}
For low frequency magnetic turbulence, i.e. in purely electromagnetic turbulence when the medium is dielectrically inactive, the dielectric function $\tilde{\vec{\in}}=\vec{\sf I}$ is the unit tensor, and the above expression reduces to 
\begin{eqnarray}
\frac{\partial}{\partial t}\bigg[\,\frac{B^2}{2\mu_0}+\frac{\epsilon_0E^2}{2} &+&\!\!\!\!\int_{-\infty}^t\!\!\!\!\!\!\!\!\mathrm{d}t'\ \vec{E}(t')\ \vec{\cdot\ \tilde{\sigma}}(t')\ \vec{\cdot\ E}(t')\, \bigg]=\nonumber \\[-1.5ex]
&&\\[-1.5ex]
&=&-\frac{1}{\mu_0}\vec{\nabla\cdot\Big(E\times B\Big)}\nonumber
\end{eqnarray}
{This second term in the parentheses on the left} becomes important only for radiation when displacement currents have to be included which in magnetic turbulence is not the case. Thus it can be dropped because it is relativistically small describing high frequency electromagnetic fluctuations in free space. {In a plasmas, because the dielectric response to fluctuations is included in the general definition of the conductivity, this reduces} the Poynting theorem to
\begin{eqnarray}
\frac{\partial}{\partial t}\bigg[\frac{B^2}{2\mu_0}&+&\!\!\!\!\int_{-\infty}^t\!\!\!\!\!\!\!\!\!\!\!\!\mathrm{d}t'\ \vec{E}(t')\ \vec{\cdot\ \tilde{\sigma}} (t')\:\: \vec{\cdot\  E}(t') \bigg]\!= \nonumber\\[-1.5ex]
&&\\[-1.5ex]
&=&-\frac{1}{\mu_0}\vec{\nabla\cdot\Big(E\times B\Big)}\nonumber
\end{eqnarray}
Equivalently, one may also define an equivalent dielectric response tensor 
\begin{equation}
\vec{\in}(t)=-\frac{2}{\epsilon_0}\int\limits_{-\infty}^t\mathrm{d}t'\vec{\sigma}(t')
\end{equation}
with Fourier transform in time
\begin{equation}\label{eq-epsilon}
\vec{\in}_\omega=\frac{2i}{\omega\epsilon_0}\vec{\sigma}_\omega
\end{equation}
Since at this place it does not bring any advantage, we stay with the former version. {We will, however, later return to it when proposing  a turbulent dispersion relation.}

The term on the right hand side is the electromagetic energy flux. It must be discussed separately for the special cases considered below. 

{For justification of the procedure in obtaining the above form and its limitations see the Appendix. Here we point out that the equation as written above seems to be independent of space $\vec{x}(t)$. This is not the case. It only does not contain any spatial derivatives on the left though they are present on the right. In order to treat it one still must take care of the explict time and implicit spatial dependence.}

{At this point, in order to avoid misunderstanding, one should take care of what is meant by energy conservation. The above equation still contains the unspecified turbulent conductivity which contains the full mechanics of turbulence that contributes to the generation of electromagnetic fluctuations. Hence total energy conservation is not meant here. The resolution of the above expression will merely relate the electromagnetic energy to the conductivity which, in order to obtain total energy conservation should subsequently be expressed through the full mechanical part of turbulence, a step which is not included in our approach. We just aim on finding the relation between the electromagnetic field fluctuations and the conductivity.}

In the following we restrict to the two cases of parallel and perpendicular propagation mentioned before. We first treat propagation of the turbulent fluctuations along the ambient magnetic field with all turbulent wave vectors $\vec{k} || \bar{\vec{B}}$. This reduces the product 
\begin{equation}
\vec{E\ \cdot\ {\tilde\sigma}\ \cdot\ E}= \sigma^TE^2
\end{equation}
to a scalar product of the squared component of the transverse electric field $E^2(t, \vec{x})$ and the transverse turbulent conductivity $\sigma^T(t, \vec{x})$, both functions of time and space. 

{A similar form of simplified Poynting theorem holds as well for perpendicular propagation $\vec{k} \perp \bar{\vec{B}}$, $\vec{E} || \bar{\vec B}$ and will be made use of farther below. In this case one defines the parallel turbulent conductivity $\sigma^{||}(t, \vec{x})$ instead of the above transverse and formulates the theory in terms of it.\footnote{{Perpendicular propagation $\vec{k\cdot\bar{B}}=0$ with $\vec{B} || \bar{\vec{B}}$, i.e. the \emph{compressive} part of the magnetic turbulence, requires a different approach. Any measurement of the full magnetic field $\bar{\vec{B}}+\vec{B}$ in all its three components allows for the easy separation of this case. One simply determines the mean magnetic field $\bar{\vec{B}}$ and separates out the magnetic fluctuations $\vec{B}_\|$. These belong to the excluded perpendicularly propagating compressive component of magnetic turbulence.}} The integral term containing the product of turbulent conductivity and electric field can then be expressed as an integral over the squared electric field amplitude times the transverse conductivity. We develop the theory for parallel propagation and subsequently write it also for the particular perpendicular case.} 

{In practice one is interested in \emph{stationary} turbulence.\footnote{{The transverse electric and magnetic energy densities may depend on the azimuthal angle $\phi$ in the plane perpendicular to $\bar{\vec{B}}$. They are understood as angular averages $(E^2,B^2)\to (2\pi)^{-1} \int_0^{2\pi}\mathrm{d}\phi \big(E^2(\phi),B^2(\phi)\big)$.}}  Then the temporal  variations depend on time differences only, and the time integral} can be brought into the form of a convolution integral
\begin{eqnarray}\label{eq-energy}
\frac{\partial}{\partial t}\bigg[\frac{B_\perp^2}{2\mu_0}&+&\!\!\!\!\int_{-\infty}^t\!\!\!\!\!\!\!\!\!\!\!\!\mathrm{d}t'\ {E_\perp}^2(t')\ {\sigma}^T(t-t') \bigg]\!\nonumber \\[-1.5ex]
&&\\[-1.5ex]
&=& -\frac{1}{\mu_0}\nabla\ \vec{\cdot}\ \Big(\vec{E_\perp\times B_\perp}\Big)\nonumber
\end{eqnarray}
{Since all quantities including the transverse turbulent conductivity depend on time and space, it is understood that for instance
\begin{equation}
\sigma^T\;\Big(t-t,' \vec{x}(t-t')\Big)
\end{equation}
where $\vec{x}(t-t')$ is the spatial coordinate. In streaming plasma it is the coordinate in the comoving frame (assuming nonrelativistic mean speeds).}

One frequently also refers to homogeneous turbulence. This is achieved when all quantities depend only on spatial distances $\vec{x-x}'$, and it is unimportant at which point in space measurements of the fluctuations have been performed. We will as well make use of this assumption below assuming that the solar wind turbulence is homogeneous. This also applies to a medium that is streaming with mean speed $\bar{\vec{v}}$. However, in a radially expanding flow like the solar wind or stellar winds, homogeneity applies only locally to spherical surfaces of constant radius $r$ and their surrounding, being two-dimensional in the angular coordinates $\theta-\theta', \phi-\phi'$. In the radial direction homogeneity implies that the total turbulent radial energy flux that crosses a spherical shell is constant. Thus the total turbulent magnetic energy density should vary as $B^2\propto (r-r')^{-1}$. This is different from the radial variation of the mean field $\bar{B}\propto (r-r')^{-2}$ which follows from mean magnetic flux conservation.  However, this is an idealization because it does not take into account the adiabatic cooling of the turbulent flow which is difficult to infer because the turbulent energy is distributed over a large number of scales. Thus homogeneity of turbulence in an expanding flow like the solar wind must be taken with caution.

\section{Link to measured power spectral density}
{Poynting's theorem of fluctuations in any of the above representations, in particular in the reduced form of Eq. (\ref{eq-energy}), refers to fluctuations in the time domain. This differs from conventional philosophy of turbulence which refers to spatial rather than temporal fluctuations. Transformed into Fourier space this is the wavenumber spectrum rather than the frequency spectrum.  Instead, observations of (electro)magnetic turbulence, e.g. in space, usually refer to measured power spectral density of the turbulent (electro)magnetic fluctuations in a broad range of} frequencies, {while seeking arguments to relate them to power spectral densities in wavenumber (spatial scales). This poses the question whether such observations can be used in order to obtain direct information about the turbulent electromagnetic response function.} 

In the following we {attempt a partial answer to this question}. It will differ from the conventional attempts of inferring the slope of the spectra and trying to construct models which reproduce those slopes. {Spectral slopes are extraordinarily sensitive to differences in the underlying physical setting which produces them.} Our aim is instead to construct the response function or its frequency spectrum and by retransformation into configuration and time domains obtain information about the response of the plasma. 

Taking the Fourier transform with respect to time of Eq. (\ref{eq-energy}) yields
\begin{equation}\label{eq-non}
\frac{(B_\perp^2)_\omega}{2\mu_0}=-\frac{i}{\omega}(E_\perp^2\sigma^T)_\omega-\frac{i}{\omega\mu_0}\nabla\ \vec{\cdot}\ \Big(\vec{E_\perp\times B_\perp}\Big)_\omega
\end{equation}
with index $\omega$ indicating that the corresponding expression is the Fourier transform taken in frequency space. Still all quantities depend on space. Fourier transforming in space yields the general expression
\begin{equation}\label{general}
\frac{(B_\perp^2)_{\omega\vec{k}}}{\mu_0}=-\frac{i}{\omega}(E_\perp^2\sigma^T)_{\omega\vec{k}}-\frac{i}{\omega\mu_0}
\Big[ \nabla\  \vec{\cdot}\: \Big(\vec{E_\perp\times B_\perp}\Big)\Big]_{\omega\vec{k}}
\end{equation}

It must be stressed that the Fourier transform of the nonlinear quantities is not a product of Fourier transforms. One has in the time domain, with $\vec{B}_\omega$ the transform of $\vec{B}(t)$,
\begin{equation}
\big(B^2(t)\big)_\omega=\frac{1}{2\pi}\int_{-\infty}^\infty \mathrm{d}\omega'\:\vec{B}_{\omega'}\ \vec{\cdot}\ \vec{B}_{\omega-\omega'}
\end{equation}
A corresponding expression holds for the electric field.

Similarly, in the space domain one has an equivalent expression for the power spectral density in wavenumber space in $d$ dimensions
\begin{equation}
\big(B^2(\vec{x})\big)_{\vec{k}}=\frac{1}{(2\pi)^d}\int_{-\infty}^\infty \mathrm{d}^dk'\:\vec{B}_{\vec{k}'}\ \vec{\cdot}\ \vec{B}_{\vec{k}-\vec{k}'}
\end{equation}
There is no obvious way of comparing the frequency space and wavenumber space spectra. In turbulence the relation between frequency and wavenumber is not known. It is not given by any dispersion relation $\omega(\vec{k})$ such that the frequency cannot be replaced by wavenumbers. Replacing the frequency element d$\omega$ by the $k$-space volume element d$^dk'$ exploiting the group velocity $\partial\omega(\vec{k})/\partial\vec{k}$ is inhibited for the same reason. {This provides a fundamental difficulty in interpreting power densities in frequency space as power densities in wavenumber space.}

\subsection{{Stationary homogeneous turbulence}}
Usually magnetic turbulence for instance in the solar wind, anywhere in space and astrophysics or in plasmas in general is dealing with kind of stationary turbulence. Under these conditions the turbulent response function depends only on the time difference $\tau=t-t'$. We also make the assumption that locally within a sufficiently extended volume the turbulence is also homogeneous depending only on spatial differences $\vec{\xi}=\vec{x-x'}$. This naturally restricts the spatial scales of the fluctuations to the linear extension of this volume in which the observtions are preformed, for instance in the radial direction in an expanding and flowing solar wind. Hence the right hand side of the third last above expression is the Fourier transform of a convolution in time and space which can be written as the product of two Fourier transforms
\begin{equation}
\frac{(B_\perp^2)_{\omega\vec{k}_\|}}{\mu_0} = -\frac{i}{\omega}\big(\sigma^T\big)_{\omega\vec{k}_\|}\big(E_\perp^2\big)_{\omega\vec{k}_\|}+\frac{k_\|}{\omega\mu_0}\big({E_\perp B_\perp}\big)_{\omega\vec{k}_\|}
\end{equation}
where, for parallel propagation, only $k_\|$ enters and the two fields, being both in the plane perpendicular to the stationary (mean plus external) field are, in addition, strictly perpendicular to each other. Since for the electromagnetic field fluctuations we have the relation of transversality 
\begin{equation}
E_{\perp k'_\|}=(\omega'/k'_\|)B_{\perp \omega'k'_\|}
\end{equation}
the last mixed term can, in stationary homogeneous turbulence, also be expressed solely through the magnetic fluctuations yielding
\begin{equation}
\frac{(B_\perp^2)_{\omega\vec{k}_\|}}{2\mu_0} = -\frac{i}{\omega}\big(\sigma^T\big)_{\omega\vec{k}_\|}\big(E_\perp^2\big)_{\omega\vec{k}_\|}+\frac{\big(B_\perp^2\big)_{\omega\vec{k}_\|}}{\mu_0}
\end{equation}

The fields in this equation are considered to be known from the observations, at least known in a certain range of frequencies where the turbulent electromagnetic field power has been measured in a region of space. If on the right the power spectral density of the electric field would as well become measured in the same frequency range{\footnote{{Measurements of this kind became available \citep{bale2005,chen2011} recently and will be referred to below.}}}, then one could form the ratio
\begin{equation}\label{eq-ratio}
\frac{(B_\perp^2)_{\omega\vec{k}_\|}}{(E_\perp^2)_{\omega\vec{k}_\|}}= -\frac{2i\mu_0}{\omega}\big(\sigma^T\big)_{\omega\vec{k}_\|}
\end{equation}
in order to obtain direct information about the spatio-temporal Fourier transform of the {transverse} turbulent response function of the plasma for turbulence propagating parallel to the mean field 
\begin{equation}\label{eq-sig-trans}
\sigma^T_{\omega\vec{k}_\|}=\frac{i\omega}{2\mu_0}\frac{(B^2_\perp)_{\omega\vec{k}_\|}}{(E_\perp^2)_{\omega\vec{k}_\|}}
\end{equation}
Here the electric and magnetic field fluctuations are both perpendicular to the mean field. {A similar formula follows if the electric fluctuations are parallel to the mean field and propagation is perpendicular. It determines the parallel turbulent response of the plasma
\begin{equation}\label{eq-sig-par}
\sigma^\|_{\omega\vec{k}_\perp}=\frac{i\omega}{2\mu_0}\frac{(B^2_\perp)_{\omega\vec{k}_\perp}}{(E_\|^2)_{\omega\vec{k}_\perp}}
\end{equation}
}

{We repeat again that these two parallel and perpendicular components of the conductivity spectrum are not the spectrum of the full turbulent conductivity. The compressive turbulent contribution is still missing. However, since observationally the compressive part can easily be separated out (see Footnote 2) when measuring the full magnetic and electric fluctuation fields, the above expressions hold for the remaining non-compressive components. Investigation of the compressive component requires a different approach. }

\subsection{{Non-homogeneous turbulence}}
The last two expressions are very useful if the turbulence is both stationary and homogeneous. Stationarity is not a big problem. Homogeneity, on the other hand, usually breaks down in streaming plasmas like the solar wind. Here one can assume stationary homogeneous turbulence to be realized only when considering a sufficiently small volume around a particular radial shell which cuts the spectrum off at a scale well below the mean field scale. If the condition of homogeneity is strongly violated then one must return to the former non-factorized expression (\ref{general}) which inhibits direct relation to observations. 

Non-homogeneous turbulence is more difficult to treat. One needs to return to Eqs. (\ref{eq-non},\ref{general}) and perform measurements of the left-hand side over an extended volume. Its spatial Fourier transform gives the combined frequency-wavenumber spectrum Eq. (\ref{general}) on the right of the electric power and response function. Its resolution requires additional assumptions which make use of the separate measurement of the electric power spectrum. It also requires consideration of the thermodynamic properties of the expansion of flow.

\subsection{{Additional remarks}}
The inverse transformation of either of the two above expressions into the time domain reads
\begin{eqnarray*}
\sigma^{T,\|}(t&-&t',\vec{x-x'})= -\frac{1}{(2\pi)^{d+1} i\mu_0}\times \\
&\times&\int\limits_{-\infty}^{+\infty}\frac{\omega\,(B^2_\perp)_{\omega\vec{k}}}{(E^2_{\perp,\|})_{\omega\vec{k}}}\mathrm{e}^{-i\omega (t-t')+i\vec{k\cdot(x-x')}}~{\mathrm{d}\omega}\mathrm{d}^dk
\end{eqnarray*}
This is, however, of lesser interest than the frequency and wavenumber dependence of the response function which contains the wanted information about the turbulent conductivity and dissipation. It should be stressed that this is not a linear response function in the sense of wave theory. It is a collisionless nonlinear response generated in the turbulence. It thus contains all the contributions of the turbulence. As a quantity obtained from measurements/observations its dependence on the velocity field is implicit to the field fluctuations.

{Eq. (\ref{eq-sig-trans}) is the desired expression in stationary homogeneous parallel turbulence. It in principle, contains all the information about the cause of magnetic turbulence in parallel propagation. Eq. (\ref{eq-sig-par}) is the corresponding expression which holds for the particular case of perpendicular propagation $\vec{E}\|\bar{\vec{B}}$ with the exclusion of  compressible turbulence where the magnetic fluctuation field is parallel to the mean field.} In stationary non-homogeneous turbulence these expressions just provide information about the immediate surrounding environment of the observation site. For more information one needs to have widely distributed observations or otherwise find a method of inverting the equation for the fields and solve the inverse problem, which is probably hardly possible.

The frequency spectra of the dissipative turbulent response function at any arbitrary location in space,  Eqs. (\ref{eq-sig-trans},\ref{eq-sig-par}) suggest that in addition to the magnetic power spectral density it is highly desirable to measure the electric spectral power density of the turbulence both in frequency and wavenumber space. Only the ratio of both provides the  turbulent response function. Once this has been obtained it should become subject to further investigation of its origin in mechanical turbulence, generation of vortices and the spectrum of electric current fluctuations. 

\subsection{Turbulent ``dispersion relation''}
{The spectrum of conductivity fluctuations determined from spectral energy densities suggests its use in a general form of a ``turbulent dispersion relation''. In turbulence such a dispersion relation, if it exists, is highly nonlinear, because the relation between linear wave modes and well-developed turbulence is washed out by cross-scale interactions. Nevertheless, in magnetic turbulence we are dealing with a broad spectrum of low frequency large-scale electromagnetic waves which, however, by no means are solutions of a dispersion relation derived from the linearized dynamic equations. The most general dispersion relation of such waves will, however, still be of the general form of electromagnetic waves propagating in any arbitrary medium
\begin{equation}
\mathcal{N}^2=\vec{{\in}}(\omega,\vec{k})
\end{equation}
with $\vec{\in}(\omega,\vec{k})$ the turbulent response function, in our case of magnetic turbulence. The latter is a highly nonlinear function resulting from  (\ref{eq-epsilon}), the turbulent response tensor $\vec{\in}(\omega,\vec{k})$. }

{If the latter can be determined from observations in dependence on frequency and wavenumber, this dispersion relation describes the nonlinear turbulent electromagnetic wave spectrum consisting of all the turbulent mutually interacting modes of which only a subset could be described as ordinary linear plasma modes. Unfortunately there is no way of its determination unless the measurement of the spatial scale spectrum in a large range of scales from the largest MHD down to electron scales would be possible. This is illusive. It is possible only under some severe assumptions like imposing the so-called ``Taylor hypothesis'' in a streaming plasma like the solar wind which is but the most simple assumption that the turbulent structures become transported by the flow across the observing instrument (in the solar wind the spacecraft). Though the approximate validity of this assumption has been demonstrated in fast solar wind flows, it is clear that it holds only for streaming velocities \emph{substantially larger} than the group velocities of the turbulent eddies. In addition it depends strongly on the angle between the streaming velocity and the local turbulent speeds of the eddies. It does \emph{not apply} in the direction perpendicular to the flow, which in the absence of other means (for instance multi-spacecraft compounds) inhibits the determination of the spatial spectrum by reference to the Taylor hypothesis.}

With these restrictions in mind, we make use of the determination of the temporal spectrum of the turbulent conductivity in order to construct an approximate turbulent dispersion relation. The connection between $\vec{\in}_\omega$ and $\vec{\sigma}_\omega$ has been given in Eq. (\ref{eq-epsilon}). Use of this relation for the two investigated cases of parallel (\ref{eq-sig-trans}) and perpendicular (\ref{eq-sig-par}) propagation of the turbulent electromagnetic fluctuations allows writing down the two corresponding dispersion relations
\begin{equation}\label{eq-disprel}
\mathcal{N}^2_{\|,\perp}\equiv\frac{c^2k_{\|,\perp}^2}{\omega^2}= \frac{2i}{\omega\epsilon_0}\sigma_{\omega\vec{k}}^{T,\|}
\end{equation}

These are the two exact electromagnetic dispersion relations of the turbulent electromagnetic fluctuation field so far excludiing compressive modes which have to be considered separately. Once the conductivity spectra both in parallel and perpendicular directions have been determined from the observations, they  provide the relations between frequency and wavenumber of these fluctuations. In general, since these relations are nonlinear and no eigenmodes will dominate, at least not in the inertial range, these dispersion relations cannot be obtained from any linear theory.
  
Unfortunately, from currently available measurements only the frequency spectrum enters these expressions. Combination with spatial spectra obtained from application of the Taylor hypothesis is tautological and thus does not provide any new information about the dispersion of the turbulent fluctuations. It cannot be used to find the dispersion relation nor the conductivity spectrum.  Measurements are incomplete and so is the dispersion relation based only on the temporal spectrum. Completion requires, in addition, the measurement of the spatial conductivity spectra. 

If these are not provided, any turbulent frequencies $\omega(\vec{k})$ are complicated functions of the turbulent wavenumber $\vec{k}$, and the above dispersion relations become implicit functional expressions of the turbulent wavenumbers. The frequency must then be understood as a local short-wavelength average $\langle\omega(\vec{k})\rangle$, and the dispersion relation holds, if at all, only approximately in the large wavelengths/small $k$ domain. 

Nevertheless with this drawback kept in mind we, in the following, apply it to the available measurements in the solar wind.

\section{Special cases}
In this section we return to Eq. (\ref{eq-ratio}) which expresses the turbulent response, i.e. the turbulent conductivity, through the spectral power densities of the turbulent magnetic and electric fluctuation fields. Let us assume that these can indeed be measured such that the spectrum (\ref{eq-sig-trans},\ref{eq-sig-par}) of turbulent conductivities (writing it without indication of the direction of propagation)
\begin{equation}\label{eq-sig-1}
\sigma_{\omega\vec{k}}=\frac{i\omega}{2\mu_0}\frac{(B^2)_{\omega\vec{k}}}{(E^2)_{\omega\vec{k}}}
\end{equation}
can be obtained  as function of frequency, at least in a certain range of frequencies. Since we have in mind application to one particular case in which such spectra have been measured, we drop the index $\vec{k}$ in the following understanding that -- in principle -- the retained index $\omega$ means both $\omega\vec{k}$. 

What concerns measurements respectively observations there is still an open problem concerning the application. Observations usually are provided only at a particular location $\vec{x}$  where field fluctuations are analyzed in the time domain. This means that only power spectral densities in frequency space are available. In some cases from multispcacraft measurements approximate spatial spectra have also been constructed. Unfortunately for our purposes no such really independent observations are available. The following considerations, though intend to be applications to solar wind observations, therefore lack the spatial dependence as they refer only to the available frequency domain. They must, just for this reason, be considered as approximations.

\subsection{Resonances}
We first ask under which conditions the response function would become infinite $\sigma_{\omega_r}\to\infty$ at some particular $\omega_r$ (here $\omega_r$ is still understood as $\omega_r\vec{k}_r$). If this happens, it means that the plasma contains a resonance at $\omega\vec{k}=\omega_r\vec{k}_r$. This can happen only when $(B^2)_{\omega_r}\to \infty$ at $\omega_r$, because $(E^2)_\omega\neq 0$ is finite for all frequencies.  In other words we trivially recover the obvious fact that in turbulence the presence of an eigenmode implies that the spectrum contains a resonance line at eigenmode frequency $\omega_r$. This resonance line may, however, not be visible, {though \citet{bale2005}, from measurements in the solar wind, claim the -- at least occasional -- existence of such a surprisingly broad line in the prospective ion-inertial range of scales in the interval $\lambda_i>\lambda>\lambda_e$ (corresponding to frequencies $\sim 1$ Hz) between ion and electron inertial lengths. They attribute it, without proof, to the presence of kinetic Alfv\'en waves, which may be correct though no mechanism is known to support it.}\footnote{{This claim is based on single spacecraft observations (offered by Cluster 4 where one should admit that Cluster does not provide useful observations of the electric power above a few Hz and one should await MMS observations to fill the gap), measurement of frequency spectra, and application of the Taylor hypothesis in order to infer about spatial scales. Indeed, kinetic Alfv\'en wave dispersion implies scales transverse to the ambient field in the range $\lambda_e<\lambda\lesssim\lambda_i$. In turbulence, there is no obvious \emph{linear} reason for these waves to be excited, though their presence might be probable at those scales. One possibility is offered by turbulent generation of current vortices/filaments carried by magnetized electrons in this ``ion inertial/dissipation'' range, as it is conventionally called, the range of nonmagnetic ions where currents are necessarily carried by electrons.  These currents can indeed be interpreted as nonlinearly steepened kinetic Alfv\'en waves. Their linear damping is comparably weak such that they might survive for long and mimic a deviation from the general power law shape in the spectrum. {Dissipation of the current vortices probably takes violently place in \emph{spontaneous reconnection} at the shortest scales, when the thickness of the vortices drops into electron scales, {though this seems not to have been confirmed for the current vortices detected by \citet{perri2012} which also are of the scale of kinetic Alfv\'en waves. If they have simply been those, then it is understandable that dissipation was small as the current filaments have not dipped into electron scales where they reconnect. The claim that reconnection is the main process of dissipation implies that such small-scale current filaments each contribute a rather small amount only to dissipation \citep{treumann2015}. It is the entire ensemble of all small-scale reconnecting current layers whose dissipation provides the destruction of turbulence in the dissipation range, not one single current filament itself as this carries only a weak magnetic field. Kinetic Alfv\'en waves, in the intermediate range will, however, already dissipate some of the energy, mainly by their weak field-aligned electric component which accelerates electrons along the field. On the other hand, not \emph{all} turbulenct structures will dissipate by reconnection. Structures of this kind are in the first place discontinuities which, particularly in the solar wind, have been detected to be of longevity and do not dissipate violently via reconnection if at all. They do not belong to proper turbulence, however, as they divide it into regions of different properties. Discontinuities repesent \emph{turbulent domain walls}. On their scales turbulence is non-homogeneous.}}}} Visibility depends on bandwidth and frequency resolution, as well as on the nonlinear evolution/saturation/dissipation of the mode. 

{The detection of the broad emission line in the spectrum suggests that the prospective kinetic Alfv\'en waves which are believed to contribute to this line do \emph{not} dissipate their energy in this range and thus do not become depleted merging into the overall spectral shape. Rather the kinetic Alfv\'en currents, for turbulent dissipation to become effective, shrink in their scales until matching the dissipation range which happens only when scales become of the order of the electron scale. This means that the kinetic Alfv\'en current filaments must first cascade down to electron scales before dissipating (though, of course kinetic Alfv\'en waves themselves posses a field-aligned electric component which will cause dissipation, e.g. by acceleration of electrons along the magnetic field). If true, the deviation from the inertial range of Kolmogorov turbulence can be understood in this way as \emph{structure formation} in current filaments. It is, however not completely clear whether this happens or not, because the electric spectral densities referred to the Cluster spacecraft coordinates. The electric field is subject to frame dependence, and the direct comparison with the magnetic spectrum might for this reason suffer from lacking knowledge. The required transformation has for other data been done by \cite{chen2011} who do not report such a line emission which, in their case, might however be related due to insufficient extension of the spectrum to small enough scales. Their frequency range was limited to $\lesssim 0.02$ Hz by the velocity measurements.

\subsection{Power laws}
Observations frequently indicate the presence of (more or less well developed) power laws in some parts of the magnetic spectral energy densities. {Since only frequency spectra are available, we restrict to a discussion of those only while keeping in mind that a correct theory should (for the above mentioned reasons) include \emph{independently} measured $k$-spectra. Thus the following considerations in this and the following sections apply either to fixed wave numbers $k$ or to frequency spectra \emph{averaged over $k$-space}. Observations usually do not indicate to which group they belong. It is, however, expected that they take an average over spatial scales, in the sense of $\langle (B^2)_{\vec{k}}\rangle_\omega$, where $\langle\dots\rangle$ means averaging over a range in wavenumbers. The following spectral dependencies are therefore understood as being averaged over wavenumbers. To indicate this we suppress the wavenumber index $\vec{k}$. We also, for simplicity, drop the angular brackets. }

 Let us assume that the magnetic power spectral density in frequency space indeed becomes power law of power $\alpha>0$ in some frequency range $\omega_{in}<\omega<\omega_d$, with $\omega_{in},\ \omega_d$ the respective frequencies of energy injection and dissipation (or transition to a different kind of power law, a break point for instance),
\begin{equation}
(B^2)_\omega\sim C_b\omega^{-\alpha}
\end{equation}
This excludes any constant conductivity that would be independent of frequency in this range. Because assuming that $\sigma=\sigma_0=$ const implies that $\sigma_\omega=\sigma_0\delta(\omega)$ and, hence $(E^2)_\omega$ independent on $\omega$ which contradicts the above equation.  {If, on the other hand, the electric spectral power density is another power law
\begin{equation}
(E^2)_\omega\sim C_e\omega^{-\beta}
\end{equation}
which usually is not known however, then 
\begin{equation}\label{eq-sig-2}
\sigma_\omega\sim C_\sigma\omega^{-(\alpha-\beta-1)}
\end{equation}
is as well power law. Depending on $\alpha>\beta+1$ (which includes the case $\beta<0$) or $\beta>\alpha-1$ the conductivity spectrum decreases or increases with increasing frequency in this range of frequencies.} 

From these simple considerations one thus concludes that in the inertial range obeying a certain power law in frequency -- to repeat: averaged over a range of wavenumbers -- the spectrum of conductivity cannot be flat because this would correspond to a constant conductivity which must be excluded for reasonable magnetic power law spectral densities, i.e. $\sigma_\omega\neq$ const. {Since any finite conductivity indicates some kind of dissipation, magnetic power laws cannot be completely dissipationless, at least in a broad sense, which is a {very interesting and sensitive point because} magnetic power laws are typical for the inertial range. {What is the meaning of `dissipation in the inertial range'? Inertial ranges are characterized \citep{kolmogorov1941} by a constant turbulent energy flow passing from long to short scales, i.e. from low to high frequencies as well. }

If, on the other hand, the electric spectral density also obeys a power law, then the conductivity spectrum $\sigma_\omega$ is as well power law. One would expect that with increasing frequency the conductivity worsens and ultimately enters into the dissipative range. Hence, the realistic case should be $\beta<\alpha$, which implies that the power spectral density of the turbulent electric fluctuation field is \emph{flatter} than that of the magnetic fluctuations. {The possibility of an inertial range power law conductivity spectrum again suggests the presence of dissipation just in the inertial range. Depending on whether this dissipation is real or imaginary, it causes heating and production of entropy, or it reflects the presence of inductive effects of the kind of blind resistivities known from transmission lines, i.e. phase shifts in the fluctuations involved.}

\subsection{Dissipation range: exponential decay}
{In the dissipative range the assumption that we are dealing solely with electromagnetic waves breaks down. The dissipative range is necessarily dominated by electrostatic fields which are required by collisionless dissipation. Thus our theory in the restricted form does not strictly apply here anymore. When entering the dissipative range of frequencies $\omega >\omega_d$ the turbulent magnetic power density spectrum is cut off. Theoretically one expects an exponential cut-off in the nonmagnetic range of scales/frequencies. The typical \emph{spatial} scale of this decay is the skin depth $\lambda_\mathit{skin}=c/ \omega_e$. Thus, theoretically, one should not observe magnetic fluctuations of shorter scales here. Observations of the magnetic power spectra in the solar wind occasionally suggest non-exponentially modified slopes in what is apparently defined as the dissipation range \citep{alexandrova2009, sahraoui2009,sahraoui2012}. These may have different reasons. Even though the solar wind is streaming at super-alfv\'enic speed, the dissipation range of the magnetic power spectrum might be polluted, for instance, by upward streaming transport of magnetic energy at shorter scales as well as by other sources of fluctuations like generation of magnetic Weibel modes at electron inertial scales etc. For the electric power spectrum one expects much stronger differences because, here, electrostatic waves will necessarily become strongly excited and will substantially modify the electric spectrum. Comparison of electric and magnetic spectra should provide information about the scale where kinetic turbulence and electrostatic electric fields take over.}

{Ignoring the anyway badly understood modifications of the magnetic power spectra, we assume that the electromagnetic (magnetic and the corresponding electric) spectral energy densities at frequencies corresponding to the dissipative range should become exponentially cut off (what concerns the measured electric power this might be invalid because of the expected contribution of electrostatic fields for which we don't raise the question of how it could be separated, observational claims may be doubted because it is hard to believe that current observations in the solar wind would resolve the shin-dpeth range)} like
\begin{equation}
(B^2)_\omega\propto\mathrm{e}^{-\gamma_b\omega}\quad\mathrm{and}\quad (E^2)_\omega\propto\mathrm{e}^{-\gamma_e\omega},~~~~~~~\omega\gtrsim\omega_d
\end{equation}
with $\gamma_b,\gamma_e$ magnetic and electric damping rates. This yields for the conductivity spectrum
\begin{equation}\label{eq-ans}
\sigma_\omega\propto\omega\;\mathrm{e}^{-(\gamma_b-\gamma_e)\omega}, \qquad \gamma_b>\gamma_e
\end{equation}
This spectrum  for $\gamma_e>0$ decays substantially less than either of the field power spectral densities separately in the dissipative range. In the other case $\gamma_b<\gamma_e$ the conductivity would increase with frequency here. This contradicts the assumption of onset of dissipation on the fast variation scales where the turbulent fluctuations enter the ``molecular" scale and thus kinetic processes should set on and damp the turbulence transferring the turbulent energy into heat. One rather expects that in the dissipation range the conductivity should indeed become drastically reduced. Therefore the exponential decay of the magnetic fluctuation field for $\gamma_e>0$ should be stronger than that of the electric fluctuations.

Stronger spectral decay of the magnetic than the electric power spectral density in the dissipation range is also required to make the inverse transformation of the conductivity convergent in this case. Plugging in the last expression for the conductivity spectrum into the inverse transform, applying the method of steepest descent -- with factor $\omega\to\exp(\ln\,\omega)$ --, and considering the dominant contribution only, we have for the equivalent conductivity
\begin{equation}
\sigma(\tau)\approx \frac{C}{2\pi\mu_0}\frac{\mathrm{e}^{-\gamma\omega_d-i(\tau\omega_d+\frac{\pi}{2})}}{\omega_d(\gamma+i\tau)}
\end{equation}
Here we used $\gamma=\gamma_b-\gamma_e>0,\ C=C_b/C_e$, and $\omega_d$ is again the frequency where the power spectral densities enter the high-frequency dissipative range, a quite loosely defined point which is usually assumed to coincide with the high-frequency end of the inertial range -- if it exists and can be detected. Note that $\gamma$ has dimension of time. 

The magnetic power spectrum must necessarily dissipate stronger, i.e. at a faster rate than the electric power spectrum. Moreover, in this most simple exponentially damped model of turbulence corresponding to a dissipative cut-off at $\omega\geq\omega_d$ in its high-frequency range, the $\tau$-dependence of the response function, i.e. in our terminology the turbulent conductivity, by whatever process it has been generated, is oscillatory on the scale of $\omega_d$ while it decays algebraically with $\tau$ like $\sigma\propto (\tau/\gamma)^{-1}$. It also contains an imaginary (inductive) term. Thus the dynamics of the turbulence included in our model implies that with increasing lapse $\tau$ the dissipation in the dissipation range should increase as indicated by the increasing effective resistivity $\eta=1/\sigma\propto(\tau/\gamma) $. Since we assumed stationarity the dependence on $\tau=\Delta t$ is a fictive time-dependence only, in fact being an artefact of our assumed model. It just reflects the fact that the dissipation at $\omega=\omega_d$ is smallest while increasing with frequency $\omega>\omega_d$.  

\subsection{Thermal fluctuations in the dissipation range}
From a low-frequency magnetic turbulence point of view, it seems unreasonable that the electric power spectral density related to the magnetic turbulence would increase instead of decay in the dissipation range. However such an increase happens if at high frequencies the spectral power of magnetic turbulence transforms into thermal fluctuations of the electric field. This implies purely kinetic heating, {which is not tractable in our electromagnetic treatment.}

Recently we suggested \citep{treumann2015} that the {ultimate and dominant dissipation of large-scale magnetohydrodynamic turbulence when cascading down into collisionless magnetic turbulence and entering the dissipation scale range is due to spontaneous reconnection in the multitude of narrow progressively produced current filaments on the electron scale.} Such filaments become generated starting in the ion inertial regime once the turbulent scale drops below the ion gyro-scale. The currents are carried solely by electrons. Dissipation in reconnection proceeds when the currents become electron-scale filaments  \citep[cf., also,][who instead of reconnection make kinetic Alfv\'en waves and whistlers responsible for the dissipation, a quasilinear substantially less violent process]{boldyr2013}. Such small-scale current filaments have been observed \emph{in situ} in the solar wind\footnote{One may note that the small-scale current sheets which \cite{perri2012} observed, also fit the scales of kinetic Alfv\'en waves \citep{sahraoui2009,sahraoui2012}. Estimates of dissipation in such currents by reconnection, as far as possible and reliable, seem to be small. The latter is not surprising for each current filament carries only a weak current. What counts for turbulence is the volume integrated dissipation of the entire ensemble of current filaments.} \citep{perri2012} and have also been found in small-scale numerical magnetohydrodynamic simulations \citep[cf., e.g.,][where they are again interpreted as kinetic Alfv\'en wave structures, not currents]{zhdankin2014}. Dissipation in spontaneous reconnection is a purely kinetic process, in which case one has $\gamma_e<0$ leading to a {drastic increase of dissipation and decay of the conductivity spectrum with frequency. In the exponential dissipation model this may be modeled as}
\begin{equation}
\sigma_\omega\propto\mathrm{e}^{-\gamma_\sigma\omega}, \qquad \omega\gtrsim\omega_d
\end{equation}
{and applies immediately after the spectrum enters into the dissipation range. One may note that this \emph{ansatz} is independent of our electromagnetic approach to the inertial range. It just says that the turbulent conductivity spectrum decays exponentially in the dissipation range, no matter what kind of magnetic spectral slope has been detected in the inertial scale and frequency domain. In the same spirit the exponential decay is understood as a model. It might be modified by other processes which would apply to observations on non-exponential magnetic dissipation-range spectra \citep[cf., e.g.,][and others]{alexandrova2009}. So far, at least to our knowledge, no corresponding electric spectral observations are available in this range.} 

{Such observations are difficult to perform because of the non-covariant nature of the electric field as this depends on the flow velocity and its fluctuations.  It must be re-transformed into the moving frame. This requires a precise measurement of the velocity with same resolution as the field imposing high demands on the measurement of the particle momentum distribution. As previously noted  \citep{treumann2016}, such problems can be avoided on the way of injection of gyrating ion beams (different from proton beams for the purpose of distinguishing them from solar wind protons). Such gyrating beams of fixed energy provide information about both the flow velocity and the electric field fluctuations and are thus independent of the problems introduced when referring to high resolution 3d-particle detectors. }

Explosive plasma heating (resulting from spontaneous reconnection in narrow current filaments) causes a drastic reduction of the conductivity spectrum, an increase in the power spectrum of thermal electric fluctuations and resistivity. {The conductivity spectrum in the dissipation range can of course formally be re-transformation} into $\tau$-space, yielding $\sigma(\tau)$ with $\gamma=\gamma_\sigma$. The electric power spectrum, when added up over all the reconnecting current filaments, should exhibit a steep increase  due to the excitation of thermal fluctuations in the \emph{electrostatic} field {at frequencies above} the electron cyclotron frequency $\omega\gtrsim\omega_{ce}$. The conductivity evolves with lapse time $\tau$. Since we did not include any energy input, the decrease in conductivity is unbalanced. Formally we may use the former solution. The high-frequency conductivity $\sigma(\tau)$ is about constant for lapse times $\tau<\gamma$ but for $\tau>\gamma$ decreases as $\sigma(\tau)\propto \tau^{-2}$.

\section{Observed turbulent power spectra}
{Observational information about magnetic turbulence is usually mainly obtained from observations of magnetic power spectra (see the cited literature).} We repeat that the magnetic field is covariant, i.e. frame independent, while the electric field is not. We refer to magnetic and some sparse electric measurements of turbulent fluctuations in the solar wind. {Here we summarize the preliminary results obtained \citep{treumann2017} when applying the above theory to those data.}

\subsection{Solar wind magnetic power spectra}
{In a streaming plasma like the solar wind the velocity of the flow has no effect on its magnetic fluctuations. Observations generally provide spectral slopes. They also determine the spectral extension of the inertial ranges. More recently, magnetic power spectra have been extended to shorter scales believed to enter the dissipation range in the attempt to obtain the shape of the magnetic dissipation spectrum. In addition, based on Taylor's hypothesis, parts of the frequency spectra have been transformed into wavenumber spectra which in several cases confirmed the applicability of Taylor's hypothesis, i.e. when the obtained spectral wavenumber slopes agreed with those of the frequency spectra, {note however, that in the high frequency (Cluster search coil) range application becomes suspect observationally as well as theoretically.}\footnote{{We thank the referees for bringing this point to our attention.}} {At high frequencies any simple transport of turbulent vortices by the flow should break down.} Agreement was usually taken as sufficiently precise, within statistical errors. Some of the attempts also used multi-spacecraft observations to infer about the spatial spectra. For the solar wind, observed inertial ranges are dominated by Kolmogorov spectra which are claimed to be robust {\citep[cf.][for well founded arguments]{podesta2006}.} Deviations are considered small and usually are waived. This may be taken reliable or not and is not important for our purposes. It should be mentioned however that theory of deterministic chaos as also numerical simulations of idealized models of turbulence indicate that deviations in slope, even tiny, frequently result from vastly different {physical} processes. In fact, the deviations of slopes in the solar wind magnetic power spectra are by no means tiny, as the histograms of spectral indices collected by \citet[][their Fig.2]{chen2011} indicate. Moreover, the mere presence of a dissipation function implies that there will always be some dissipation, even in the inertial range. We just mention this fact. Whether or not this claim is in agreement with any dissipationless energy flow across the spectrum as required for the inertial range, we do not discuss here though it is a fundamental unresolved question in turbulence theory.}

\subsection{Solar wind electric power spectra}
{In the near past some few measurements of turbulent electric power spectra in the solar wind have been published based on one single Cluster {\citep{bale2005,sahraoui2009}} and the {\small{ARTEMIS}} spacecraft \citep{chen2011} observations, the latter consisting of two closely spaced satellites. Unfortunately, the {\small{ARTEMIS}} observations did not make use of the distance between the spacecraft. They do not provide any spatial power spectra of the turbulent electromagnetic field. This inhibits the full use of our expression while restricting us to remain with only the frequency spectrum. Moreover, these measurements are single spacecraft measurements but used a large number of time intervals of observation to obtain average electric power spectra, all obtained under slow solar wind conditions in one spatial location, i.e. mean streaming velocities of the order of $\sim500$ km\,s$^{-1}$. The electric field on Cluster was obtained from a double probe, while {\small{ARTEMIS}} measured it  with a wire-beam antenna. Since the electric field depends on the frame of measurement, observations at spacecraft and in the moving solar wind necessarily differ. In order to obtain comoving spectra spacecraft measurements must be transformed into the solar wind frame according to $\vec{E}_{sw}=\vec{E}+\vec{v}_{sw}\times\vec{B}$, which requires measurements of the velocity. This has been done in the case of {\small{ARTEMIS}}. Spacecraft-frame measurements of the electric field suggest -- within the above mentioned restrictions -- almost identical Kolmogorov slopes $\sim-\frac{5}{3}$ for magnetic and electric power spectra \citep{bale2005,chen2011} in the inertial subrange identified as linear in a log power - log frequency representation, which is somewhat surprising. Transformation to the solar wind \citep{chen2011} showed that the stream-frame speed-cleaned electric power spectra had different slope $\sim -\frac{3}{2}$. {For our application it is important to note that the latter observations have been performed in a small spatial volume in the solar wind, small enough to consider the turbulence to be about homogeneous as the  effect of radial expansion can be neglected. However the noted lack of power spectra in wavenumbers which is a necessary independent ingredient to theory is a severe drawback for the application with which we must live. Thus, though we will arrive at some tentative conclusions, it must be kept in mind that these are incomplete. The advantage of homogeneity of the turbulence is lost here by the unavailability of \emph{independently measured} wavenumber spectra.}}

\subsection{Inferred turbulent conductivity spectrum}
{What is relevant for our theory is the difference in the spectral powers in the streaming frame. Knowing both power law slopes in the streaming frame allows direct use of Eq. (\ref{eq-sig-2}). Let us for the moment ignore the large spread in both slopes and take the mean values as reliable or, as the observers say, sufficiently robust. The spectrum of turbulent conductivities in the streaming frame then becomes 
\begin{equation}
\sigma_\omega\propto \omega^{-\zeta}, \quad \mathrm{with}\quad \zeta \approx\frac{5}{3}-1-\frac{3}{2}= -\frac{5}{6}
\end{equation}
which has the same degree of robustness as the observer's claim. We should note that the use of spacecraft spectra for the electric field would not make any sense because it would be completely unphysical. Thus by this finding, in the Kolmogorov-inertial range of solar wind magnetic turbulence the turbulent conductivity spectrum increases with frequency. Apparently the conductivity becomes larger at higher frequency, which means that any dissipation decreases with decreasing scales. For the turbulent resistivity spectrum this consequently implies a decrease which {-- in the average --} is very close to a shot-noise spectrum of slope 
\begin{equation}
\eta_\omega\sim \omega^{-1} 
\end{equation}
This is indeed an interesting result. If true (within our assumptions), collisionless dissipation -- as indicated by the frequency spectrum of the conductivity -- in the inertial solar wind turbulence becomes \emph{increasingly dissipationless} with increasing frequency (and {decreasing} scale). {Higher frequencies or shorter scales imply less transformation of turbulent energy into other forms than at lower frequencies and presumably longer scales. One explanation of such a behavior is that in the inertial range a process of self-organization is acting, which progressivey generates small-scale structure here. Such a behaviour is possible in open systems, and indeed the inertial range is an open system with energy inflow at long and outflow at short scales. In the intermediate power law range it seems that the inertial range actively works to produce new structure, i.e. ever smaller-scale current filaments.}} 

{The negative exponent $\zeta<0$ in the conductivity spectrum holds only for the inertial range, which does not inhibit re-transformation into $\tau=t-t'$ temporal dependence of the turbulent conductivity by integration over the range $\omega_i<\omega<\omega_d$ only. The resulting expression can be represented by incomplete Gamma-functions of complex argument which we do not discuss at this place as it does not provide any further information.}

{It is more interesting to ask what the effect of the comparably large scatter in the power law index on the above conclusion is. We know that power law indices, the slopes of turbulent power law spectral energy densities, are rather sensitive to the underlying physics. Small deviations in slopes indicate vastly different physics. The observations provide only bounds on the magnetic power-laws 
\begin{equation}
1.5\lesssim \alpha \lesssim1.8
\end{equation}
Thus, in order to proceed, we take for the electric power spectral index the only given value $\beta=1.5$. This yields 
\begin{equation}
\sigma_\omega\sim \bigg\{
\begin{array}{ccc}
 \omega^{+1}, & \quad\quad   &  \min\{\alpha\}\ \sim \frac{3}{2} \\[-1ex]
  &   &   \\[-1ex]
 \omega^{+0.7}, & \quad\quad  &  \sup\{\alpha\}\ \sim \frac{9}{5}
\end{array}
\end{equation}
The difference is not very large though it is remarkable that in both cases the power in the conductivity spectrum is positive. This is of course true because only one fixed slope entered for the electric fluctuation power spectrum.}

{We stress that these results must be taken with caution. They are only partially valid because of the assumptions involved. Even accepting them they still suffer from the missing independently measured wavenumber spectra. It is not entirely clear what the practical consequences are of this lack of information for the interpretation of the results. Future observations filling this gap are needed. This remark even stronger applies to the next section on the turbulent dispersion relation.}

\subsection{Inferred turbulent ``dispersion relation"}
{We may use the last result on the spectra in the tentative dispersion relation derived in Eq. (\ref{eq-disprel}). This equation is useful when the frequency and wavenumber spectra are available. If only frequency spectra exist its application becomes questioned. In discussing the approximate conductivity spectrum we referred to some averaging process in wavenumber space. We obtained something like a mean conductivity spectrum which accounted for all contributions from the various wavenumbers (spatial scales) to frequency, a not unreasonable viewpoint, because a given frequency $\omega(\vec{k})$ might contain contributions from all kinds of small scale structures.} 

{Speaking about dispersion relations it is unclear, what average wavenumbers mean. In the refraction index they have a given precise meaning. This meaning is lost if  we are not in the possession of both wavenumber and frequency power spectral densities. So we should be aware when writing down what we call here a `dispersion relation' that it refers on the one hand to the wavenumber in the refraction index, on the other to some average not well defined scales contained in the averaged frequencies. In applying Eq. (\ref{eq-disprel}) the above powers of the electric and magnetic frequency spectra are understood in terms of scale-averaged frequencies $\langle\omega(k)\rangle$. Using them formally, }based on measurements in the solar wind, wavenumbers increase with frequency:
\begin{equation}
k^2\ \propto\  \langle\omega(k)\rangle^{1.8}\quad \Longrightarrow\quad k\ \propto \pm\;\langle\omega(k)\rangle^{0.92}
\end{equation}
{(Note that we have re-introduced here the angular bracket averaging notation.)} Assuming a sliding average, then $\langle\omega\rangle=\langle\omega_0\rangle+\langle\omega_0\rangle'(k-k_0)$ can be expanded with respect to $k$. {This gives}
\begin{equation}
{k\ \propto\  \pm\langle\omega_0\rangle^{0.92}\big\{1+0.92\big[\log\langle\omega_0\rangle\big]'(k-k_0)\big\}}
\end{equation}
Though being close to linear, i.e. close to the spectrum of Alfv\'en or magnetohydrodynamic waves in general, a dispersion relation of this kind has no equivalent in linear wave theory. It reflects the complicated mixing process of waves and scales in collisionless low-frequency magnetic turbulence in the particular case of the solar wind.

{The increase in { wavenumber $k$ with frequency $\langle\omega\rangle$} corresponds to the expected decrease in the turbulent wavelengths -- spatial scales -- with increasing frequency. It is therefore pointing in the right direction. What is surprising in this result is that the decrease in spatial scales is nearly inverse to frequency. Thus, what in frequency appears as a steep spectrum consists effectively of a turbulent medium of inversely decreasing wavelengths. An extended discussion of these points and their physical implications are given elsewhere \citep{treumann2017}. {Summarising here, we note that our conlcusions hold only for the inertial range and for the particular case in which measurements in the slow solar wind have been available \citep{chen2011}. If we again take care for the scatter in the slope of the magnetic power spectral densities, we find a substantial variation in the inferred dispersion relations:
\begin{equation}
\langle\omega(k)\rangle\sim \bigg\{
\begin{array}{ccc}
 k^{+4}, & \quad\quad   &  \min\{\alpha\}\ \sim \frac{3}{2} \\[-1ex]
  &   &   \\[-1ex]
 k^{+\frac{3}{2}}, & \quad\quad  &  \sup\{\alpha\}\ \sim \frac{9}{5}
\end{array}
\end{equation} 
As before, the frequency increases with wavenumber but now at a much higher rate. Though this is in agreement with the expectation that higher frequencies correspond to shorter scales in the turbulence, the increase in the marginal spectra is surprisingly steep. When seen from the pont of view of phase and group velocities of the involved fluctuations one understands what physically happens. Both, the group and phase velocities increase with scale according to
\begin{equation}
\langle v\rangle_\mathit{ph,gr}(k)\sim \bigg\{
\begin{array}{ccc}
 k^{+3}, & \quad\quad   &  \min\{\alpha\}\ \sim \frac{3}{2} \\[-1ex]
  &   &   \\[-1ex]
 k^{+\frac{1}{2}}, & \quad\quad  &  \sup\{\alpha\}\ \sim \frac{9}{5}
\end{array}
\end{equation}
Shorter scale fluctuations run away at faster speeds. They also transport energy at faster speed away than longer wavelengths. For the flatter spectra this happens at a substantially faster rate than for steep spectra. In all cases, however, including the mean dispersion relation, the run-away of the shorter scales implies that structure formation by nonlinear processes acts in the inertial range. Speculatively one might conclude that the inertial range when characterized by this kind of structure formation is the range where gradually smaller-scale current filaments form in order to prepare their subsequent dissipation in the dissipation range of turbulence. Hence, the increase in the spectrum of conductivity and apparent decrease in dissipation in this kind of an open system does not imply that dissipation (transformation into heat and entropy increase) is necessarily decreasing or increasing but that the turbulence re-organizes into finer and finer structures. It would be extraordinarily important to infer about the thermodynamics in this spectral range and the structure of the particle distribution. Observations \citep{bale2005} seem to hint on the presence of whistlers or kinetic Alfv\'en waves in this range which might be involved in structure formation. Distribution functions should indicate whether intermittency plays a role here.}}

{We note that measured wave frequencies by the four Cluster spacecraft at proton scales, not referring to the Taylor hypothesis,  were close to zero over almost a decade of wave numbers. It is not quite clear what this means. It indicates, at least, that a dispersion relation derived in the above way without available reference to wavenumber spectra might not meet reality. In order to obtain it one requires precise and independent knowledge about both the frequency and wavenumber spectra of the electric and magnetic turbulent fluctuation fields in order to construct the correct turbulent response function.}\footnote{The claimed detection of vanishing frequency $\omega\sim 0$ on all scales in turbulence is disturbing. It implies that in the frame where the velocity fluctuates $\omega'\approx \vec{k\cdot v}$, meaning that in the entire range of observations the field would to all scales be frozen to the plasma -- in fact it is the field which is frozen because the energy is contained in the plasma flow. This is hard to understand in the range outside MHD when e.g. the ions decouple and the field is frozen only to the electron gyration with electrons along the field being free. For the electric field this implies that $\delta\vec{E}'=-\delta\vec{v\times}\bar{\vec{B}}-\bar{\vec{v}}\times\delta\vec{B}$ with the second term dropped in the mean comoving frame, i.e. $\delta\vec{v}\sim\delta\vec{E}\times\bar{\vec{B}}$ for all scales, a rather severe condition that excludes all dissipative plasma processes even on the fluctuation scale. Turbulence would be entirely mechanical then with the electromagnetic field reacting completely passively to it. Probably vanishing frequency is either approximate only within the error of measurement, or it indicates that one is still in the domain of ideal MHD scales. Otherwise, if confirmed by independent observations of frequency and wavenumber spectra, it implies that there is no turbulent electromagnetic dissipation on any scales longer than the electron inertial scale. Solar wind turbulence in this case is dissipated in three ways mechanically, by formation of discontinuities/domain walls, and electromagnetically ultimately in spontaneous reconnection at the electron inertial scales.}

{It is, however, clear that these conclusions have to be taken with care because of our lack of knowledge of the spatial power  spectral densities which inhibits construction of the correct turbulent dispersion relation.}

\section{Conclusions}
{The result of this straightforward investigation of the energy transport equation in low-frequency electromagnetic turbulence is Eq. (\ref{eq-sig-1}), which is the frequency spectrum of the turbulent response function. On should keep in mind that this does not imply total energy conservation as the mechanical part of turbulence is not included explicitly! Guided by the basic equations of electrodynamics the application of the theory given here refers to temporal spectra which, in the application, should be understood as averaged over spatial scales. The theory itself provides the full wavenumber-frequency spectrum of the response function and has quite general prospects of application \emph{in case both can be independently measured}. The interpretational problem arises when trying to explain the obtained turbulent conductivity spectrum by some model of turbulence. This requires a discussion of the turbulent Ohm's law which we wrote down for two particular cases but did not investigate further. Nevertheless we were able to derive a quite general useful expression for the turbulent conductivity spectrum in frequency and wavenumber space under the restriction to stationary and homogeneous turbulence but without reference to its mechanical part. The assumptions made are surprisingly weak. Stationarity, even in the streaming solar wind, is probably well justified as otherwise even the observations would partially fail. Homogeneity is a more subtle problem in a streaming and expanding plasma. It restricts application to a particular spatial shell whose extension must be less than the typical scale of expansion of the flow. Observationally this is not very problematic. By using Poynting's theorem the nonlinearity of the fluctuations is retained. The main simplifications concerned the neglect of third-order correlations, neglect of electrostatic contributions and restriction to the mean field frame. Moreover, for tractability reasons we considered only two directions of propagation of any electromagnetic turbulent fluctuations: propagation parallel, and propagation perpendicular to the mean magnetic field, in the latter case with electric fluctuation field parallel to the mean field. This excludes the case of perpendicular propagation and magnetic fluctuations parallel to the mean field. So the theory is not completely general. It covers not all directions of oblique propagation. Within these restrictions the turbulent conductivity spectrum was obtained for these two propagation modes.}  

{Application to solar wind observations of these expressions made use of the observationally determined spectral slopes of the magnetic and electric power spectral densities in the solar wind. This was possible only in the one case of propagation, where both electric and magnetic power spectral densities were available and the electric fluctuations were correctly transformed into the proper mean moving frame of the solar wind \citep{chen2011}. In this case the wavenumber averaged frequency spectrum of the turbulent conductivity could be determined. It was found to become an approximate shot-noise spectrum,indicating the presence of violent chaotic interactions and self-organization. This spectrum also provided the possibility to derive a wavenumber averaged turbulent dispersion relation based on the observations of the turbulent solar wind power spectra.}

{The dispersion relation, though still incomplete and approximate, as no spatial spectra could be used, is a rather approximative turbulent dispersion relation. It correctly expresses the increase in wavenumber with increasing frequency. It also indicates that in the solar wind the spatial dispersion of the turbulence depends weakly though nonlinearly on frequency. One may claim from its structure that such dispersion relations cannot be obtained from any linear theory. It remains unclear, however, what effect averaging over wavenumber has.}

{Turbulence is not a conglomerate or mixture of linear modes. In Fourier space it is the formal transformation of the observed spatially and temporally structured fluctuation fields into their Fourier equivalents which cannot by any means be reduced to a superposition of linear eigenmodes. This is exhibited in the obtained dispersion relation. There is no linear eigenmode equation available in turbulence which is prevented by the mixture of mechanical motion and electromagnetism. Such a separation is, to some extent, possible in weak turbulence, where the evolution of turbulence can be represented as mode coupling with the number of modes steeply increasing with growing wave amplitude. \citet{yoon2007a} and \citet{yoon2007b} more recently developed the rigorous kinetic weak-turbulence theory of magnetic turbulence for parallel propagation at low frequencies. Only a small number of waves are involved, viz. Alfv\'en-whistler and slow magnetosonic modes including their nonlinear modifications up to third order in the expansion with respect to normalized wave power. The obtained expressions are highly involved, which makes their direct application to observations difficult.}

{Observed spectra are featureless indicating that turbulence is rather not weak: well developed featureless turbulence seems not to be ``renormalizable''. Instead one must switch to a non-perturbative approach. A magnetic turbulence theory of this kind has not yet been developed \citep[for review of some attempts cf.,][]{verma2004}. The dimensional considerations of \citet{kolmogorov1941} provided an important global step in the non-perturbative direction but dealt with the general problem of hydrodynamic turbulence, not having had in mind any application to  electromagnetic turbulence. Global treatments of magnetohydrodynamic turbulence in Kolmogorov's spirit followed much later \citep[cf.,][and others]{iroshnikov1963,kraichnan1964}.} 

\section{{Summary and critical remarks}}
{Summarising, we draw the following brief conclusions. Since observations of turbulence in the solar wind provide mainly spectral power densities of the turbulent magnetic fluctuations in the frequency domain, very rarely accompanied by those of electric fluctuations, and to our knowledge almost never included those of the velocity fields {\citep[cf.,][for preliminary observations]{podesta2006}}, we attempted a formulation of turbulence solely in terms of electrodynamic quantities. In fact, magnetic fluctuations are the easiest to measure. Electric fluctuations are not covariant and require transformation into the moving frame which in its turn requires knowledge about the velocity fluctuations. There are ways of measuring the electric fluctuations directly by antennas, double probes, and injection of electron or ion beams, with the latter not yet explored. In these cases our bold attempt is well justified. All the mechanical part is included in the response function of the plasma. We have shown that this is in principle possible when the compressive part of turbulence is separated from the noncompressive turbulence. Under some restrictions this procedure yields expressions for the turbulent response functions, which can be investigated in view of their physical content.}

{This attempt is, however, not satisfactory what concerns all of its results. As long as only the power spectral densities in the frequency domain are considered one cannot expect that physical results are obtained which can be trusted. In order to proceed one} {absolutely requires both \emph{independently} measured power spectral densities of the electric and magnetic fluctuations in frequency and wavenumber space. It is most important that these measurements are {independent} not making use of any Taylor hypothesis or similar relations. Only the combination of both kinds of observations will reproduce the correct physically reasonable response function\footnote{But see the former footnote!}. These two kinds of spectra must be combined in the calculation of the turbulent dissipative response function. It is insufficient to use wavenumber spectra to transform them into frequency spectra via the Taylor approach, which anyway becomes suspect at high frequencies. Both must independently be used in constructing the response function. Moreover, a way has to be found for independently transform the electric spectra into the moving frame in order to make them Lorentz invariant. Tranformation to the mean-field frame is insufficient even as a first step for making the field approximately covariant. Also the field fluctuations need to be transformed. Thus other experimental ways have to be found for determining the electric field in various frequency and wavenumber ranges.} 

The results we presented can only be considered as approximate for the above reasons. We have constructed a (not completely covariant) wavenumber-space averaged response function which using observations obeyed a power law. 

We also have shown in which way an electromagnetic turbulent dispersion relation can be obtained. The dispersion relation we provide makes little sense just for the reason that it is based on averaging over wavenumber space.  It just demonstrates how one would have to construct the dispersion relation. 

A dispersion relation in turbulence does not mean that real modes are detected. It means that there is a nonlinear relation between wave numbers and frequency which is obtained by Fourier transformation in space and time. It does not result from eigenmode equations. It, however, contains inportant information about the relation between the spatial and temporal structure of the turbulence in different scale ranges.

{Future observations should consider these lacks, repair the deficiencies, fill the gaps, note the caveats and try to provide complete and reliable data. In particular necessary are independent spectra in frequency and wavenumber, measurements of electric spectra, observation of covariance and either measurements of velocity fluctuations or circumvent them by appropriate measuring techniques. Velocity spectra are very difficult to obtain independently. They require particle detectors of very high angular, energy and time resolution. It would be important of combining electric drift field observations with measurements of electric fields by other means (double probes, antennas, injection of ion beams which, in particular, have a number of advantages over electron beams: easy identification, greater stability, covering frequencies below electron cyclotron) in order to infer about the Lorentz term $\vec{v}\times\vec{B}$ as function of all temporal and spatial scales.} 

{This is still musics of the future. We nevertheless believe that proceeding in the indicated direction should be a promising way in the investigation of solar wind turbulence outside the kinetic frequency and wavenumber scale ranges. It would also contribute to the general understanding of turbulence.}

\begin{acknowledgement}
This work was part of a Visiting Scientist Programme at the International Space Science Institute Bern executed by RT in 2007. We acknowledge the rather reserved interest of the ISSI Directorate and the friendly hospitality of the ISSI staff. We thank the ISSI system administrator S. Saliba for technical support and the librarians Andrea Fischer and Irmela Schweizer for access to the library and literature. {We also acknowledge the very valuable critical comments of the referees, indicating the merits and limitations of spacecraft observations in the solar wind and, in particular, bringing the ARTEMIS data to our attention.}
\end{acknowledgement}

\section*{{Appendix: Averaging procedure, simplifying assumptions, approximations, caveats}}
{The present approach is based on averaging the Eq. (\ref{eq-poynting}). For transparency it is necessary to go into more detail. Taking the average of some quantity like the magnetic field $\vec{B}$ implies performing temporal and spatial integrations. The magnetic field is defined as magnetic flux per cross section which implies that a spatial volume is involved. Moreover, the field varies in time. Being interested in comparably short spatial and temporal scales, the scales on which the field fluctuates, not the larger scales of some mean field behavior, we assume that $\vec{B}(\xi,\tau)$ depends on two spatial scales $L\gg\ell$ such that $\xi(\vec{X},\vec{x})$ is a function of the large-scale coordinate $\vec{X}$ which varies on scale $L$, and the small-scale coordinate $\vec{x}$ which varies on scale $\ell$. Similarly, $\tau(T,t)$ depends on the slow time $T$ and fast time $t$. In order to be able to distinguish between these two different scales one integrates over the short spatial and fast temporal scales to obtain the mean-field quantity
\begin{equation}
\bar{\vec{B}}\equiv\big\langle\vec{B}(\vec{X},T)\big\rangle = \frac{1}{\theta V_x}\int_0^{\theta}\mathrm{d}t\int\mathrm{d}x^3\vec{B}(\xi,\tau)
\end{equation}
where $\theta$ is the longest fluctuation time scale and $V_x\sim \Lambda^3$ the fluctuation volume, with $\Lambda$ the longest fluctuation scale. Having obtained the mean field in this way we assume that the total field $\vec{B}=\bar{\vec{B}}(\vec{X},T)+\delta\vec{B}(\vec{x},t)$ is decomposed into the superposition of mean and fluctuating fields. This is the usual way of distinguishing between both components. Clearly, the average of the fluctuations $\big\langle\delta\vec{B}\big\rangle=0$ taken in the above sense vanishes, while in general higher order products do not necessarily vanish.}

{We apply this procedure to Eq. (\ref{eq-poynting}) where the second term on the left will be suppressed from the beginning because it is relativistically small. In the description of turbulence, radiation is excluded as all fluctuations are confined to the plasma. This is, however, not true in the dissipation range where a remarkable fraction of energy may become transformed into radiation losses. This is known from radiative cooling effects in closed astrophysical systems, clusters of galaxies, for instance.}

{The decomposition of $\vec{B}=\bar{\vec{B}}(T)+\delta\vec{B}(t)$ implies that $B^2=\bar{{B}}^2+2\bar{\vec{B}}\cdot\delta\vec{B}+\delta{B}^2$. Averaging over the short timescale the second term vanishes which leaves the slow-time derivative of the sum $\bar{B}^2+\ \overline{\delta{B}^2}$ with both terms depending on $T$. By the same reasoning averaging the right-hand terms one obtains the mean field equation
\begin{eqnarray}
\frac{1}{2\mu_0}\frac{\partial}{\partial T}\Big(\bar{B}^2+\overline{\delta\vec{B}^2}\Big)&=& -\bar{\vec{J}}\cdot\bar{\vec{E'}}-
\overline{\delta\vec{J}\cdot\delta\vec{E'}} +\nonumber\\&+&\frac{1}{\mu_0}\nabla_{\vec{X}}\cdot\Big(\bar{\vec{E'}}\times\bar{\vec{B}}+\overline{\delta\vec{E'}\times\delta\vec{B}}\Big)\nonumber
\end{eqnarray}
in which, if one would be interested in mean-field evolution (for instance in dynamo theories etc.) one should express the primed electric field, mean and fluctuating currents, and the average products accordingly. }

{Our interest is in fluctuations. So we subtract the mean field equation from the full Eq. (\ref{eq-poynting}). The latter includes the mean and fluctuating fields. This leaves us with all the remaining fluctuating terms and will turn out to be very complicated such that we need to introduce additional simplifications. Before simplifying we have:
\begin{eqnarray*}
\frac{1}{2\mu_0}\frac{\partial}{\partial t}\Big(\delta\vec{B}^2 &+& 2\bar{\vec{B}}\cdot\delta\vec{B}-\overline{\delta B^2}\Big) = -\delta\vec{J}\cdot\bar{\vec{E'}}-\bar{\vec{J}}\cdot\delta\vec{E'}-\nonumber \\
&-&\delta\vec{J}\cdot\delta\vec{E'}+\overline{\delta\vec{J}\cdot\vec{E'}}+\frac{1}{\mu_0}\nabla\times\Big(\delta\vec{E'}\times\bar{\vec{B}}+\nonumber\\
&+&\bar{\vec{E'}}\times\delta\vec{B}+\delta\vec{E'}\times\delta\vec{B}-\overline{\delta\vec{E'}\times\delta\vec{B}}\Big)
\end{eqnarray*}
All averages which are not multiplied by fluctuations depend only on the mean scales. They are constants on the fluctuation scales and, assuming that the boundaries are shifted to infinity, play no role. They may be dropped. We also assume that the mean electric field can be put to zero. This might be a severe condition but is imposed for reasons of simplicity. Moreover, we have found that the compressive magnetic fluctuation fields with $\delta\vec{B}\|\bar{\vec{B}}$ along the mean plus (external field) must be treated differently. Since observationally it is completely unproblematic to separate this component out, we put it to zero. These well justified simplifications leave us with
\begin{displaymath}
\frac{1}{\mu_0}\frac{\partial(\delta{B}^2)}{\partial t} = -\bar{\vec{J}}\cdot\delta\vec{E'}\ -\ 
\delta\vec{J}\cdot\delta\vec{E'}+\frac{1}{\mu_0}\nabla\times\Big(\delta\vec{E'}\times\bar{\vec{B}}
\Big)
\end{displaymath}
Since we are dealing with purely electromagnetic fluctuations only the very last non-averaged term in the Poynting flux on the right has been transferred to the left where it adds to the time derivative of the squared fluctuations. }  

{Now, we must consider the current. By the general Ohm's law we have $\vec{J}=\vec{\sigma}\cdot\vec{E'}$ where $\vec{\sigma}(\xi,\tau)$ is a completely general conductivity of unknown and unspecified functional tensorial form. It depends on the large and small scale coordinates and, possibly, also on the fields. Its dependencies are left open here. Formally, however, we may write it as the sum of large and small scale conductivities $\vec{\sigma}=\bar{\vec{\sigma}}+\delta\vec{\sigma}$ assuming that this would be manageable in some way. Then we have for the average current within the above made assumptions 
\begin{displaymath}
\bar{\vec{J}}=\bar{\vec{\sigma}}\cdot\bar{\vec{E'}} +\ \overline{\delta\vec{\sigma}\cdot\delta\vec{E'}}
\end{displaymath}
Again, with the same assumptions we have for the fluctuating current
\begin{displaymath}
\delta\vec{J}=\delta\vec{\sigma}\cdot\bar{\vec{E'}}+ \bar{\vec{\sigma}}\cdot\delta\vec{E'} + \delta\vec{\sigma}\cdot\delta\vec{E'}
\end{displaymath}
The first term on the right drops out because of our assumptions. The second and third terms can be combined into one when introducing back our definition of the general conductivity tensor of whatsoever form he might be. So we obtain
\begin{displaymath}
\delta\vec{J}=\vec{\sigma}\cdot\delta\vec{E'} 
\end{displaymath}
With these approximations we have 
\begin{displaymath}
\frac{1}{\mu_0}\frac{\partial(\delta{B}^2)}{\partial t} =-\delta\vec{E'}\cdot\vec{\sigma}\cdot\delta\vec{E'}\ -\frac{1}{\mu_0}\nabla\cdot\Big(\delta\vec{E'}\times\bar{\vec{B}}
\Big)
\end{displaymath}
In order to confirm to our approach the last term must vanish. This is achieved in two cases, either $\delta\vec{E'}\|\bar{\vec{B}}$, when the primed electric field component is along the mean magnetic field, or when $\vec{k}\times\delta\vec{E'}\times\bar{\vec{B}}=0$. This, for the primed electric field not parallel to the mean magnetic field implies that $\vec{k}\|\bar{\vec{B}}$. So propagation must be along the mean field. }

{In the case $\vec{E'}\|\bar{\vec{B}}$ the fluctuation of the primed electric field is given by 
\begin{displaymath}
\delta\vec{E'}=\delta\vec{E}+\delta\vec{v}\times\bar{\vec{B}} 
\end{displaymath}
where\ \ $\bar{\vec{v}}=0$\ \ is used as before and \ higher\ correlations are dropped. In the presence of finite conductivity, the primed electric field fluctuation will not vanish. This, in addition to the neglect of all higher correlations which in turbulence theory naturally play a role, in particular when the amplitudes of the fluctuations grow, is a weak point of our theory.} 

{Any velocity fluctuations in the moving frame must be perpendicular to both the mean magnetic field and the electric field fluctuations for application to observations, even when transforming into the average moving frame: $\delta\vec{v}=\delta\vec{E}\times\bar{\vec{B}}/\bar{B}^2$, which, for instance, is typical for Alfv\'en waves but in other wave modes is not generally the case. Thus neglect of the last term in the fluctuating Poynting theorem is as well problematic.} 

{This is  the most problematic of our various assumptions. It is also equivalent to not considering mean-field aligned velocity fluctuations. One would wish to improve on the theory by extending it to include higher correlations in the spirit of a perturbation expansion where the present result is understood as a first order approximation.}

{With the last term discussed away, which holds in particular cases, the final version of Poynting's law has the remarkable property of not containing any spatial derivatives. All spatial dependence is implicit. It can thus be treated applying Fourier tansformations in both space and time. This is done in the main text for the two particular cases mentioned above.}
\newpage
{It should be noted at this occasion that the complete spectrum implies the combination of frequency and wave number spectra. Considering  only one of them implies lack of half of the information of turbulence. This is the very important point to insist on as a main warning!} 

{Note that this warning is completely independent of the validity of the celebrated Taylor-hypothesis. The latter applies solely to the transport of structures across one particular spatial location by the mean flow and not to turbulence. It has no other deeper physical meaning. It is just a transformation, not even a relativistic one as inclusion of relativistic flows -- for instance astrophysical flows -- do modify it substantially.} 

{What concerns observations of electromagnetic spectral power densities and conclusions drawn from them on the nature of turbulence, we note that measuring solely frequency or wavenumber spectra will never provide any complete information.} 

{This is an important point to keep in mind in any application of turbulence theory. In our case where we are not in the position to have both at hand in application to the solar wind, the results must therefore be taken with great care. They are just preliminary in view of what really happens in the solar wind when it becomes turbulent, even though we have been in the lucky position to refer to both frequency spectra of electric and magnetic power \citep{chen2011}.}

{Any conclusions drawn on turbulence published in the literature and being based on one or the other power spectrum whithout combining them, transforming the electric spectra into the moving frame and, in addition, combining them with spatial spectra are of very little physical use as they do not provide any deeper insight into the physics of the turbulence. 
}
\end{document}